%% file: main.tex
\documentclass[11pt,a4paper]{article}
\usepackage[utf8] {inputenc}
\usepackage{amsmath,amsfonts,amssymb}
\usepackage{graphicx}
\usepackage{psfrag}
\usepackage{multicol}

\usepackage{lmodern}
\usepackage{empheq}
\usepackage{auto-pst-pdf}

\usepackage{authblk}
\title{ \textbf{Plurality Consensus in the Gossip Model} }

\author[1]{L. Becchetti}
\author[2]{A. Clementi}
\author[1]{E. Natale}
\author[1]{F. Pasquale}
\author[1]{R. Silvestri}

\affil[1]{\emph{Sapienza} Universit\`a di Roma, {\tt becchett@dis.uniroma1.it}, {\tt natale@di.uniroma1.it}, {\tt pasquale@di.uniroma1.it},  {\tt silvestri@di.uniroma1.it}   }
 
\affil[2]{Universit\`a \emph{Tor Vergata} di Roma, {\tt clementi@mat.uniroma2.it}}

\newtheorem{definition}{Definition}

\newtheorem{lemma}[definition]{Lemma}
\newtheorem{claim}{Claim}
\newtheorem{theorem}[definition]{Theorem}
\newtheorem{cor}[definition]{Corollary}

\newcommand{\Prob}[2]{\mathbf{P}_{#1} \left( #2 \right)}
\newcommand{\probb}[1]{\mathbf{P}\left( #1 \right)}
\newcommand{\probc}[2]{\mathbf{P}\left(\left. #1 \;\right| #2 \right)}
\newcommand{\expe}[1]{\mathbf{E}\left[#1\right]}
\newcommand{\expec}[2]{\mathbf{E}\left[\left. #1 \;\right| #2 \right]}

\newcommand{\proof}{\noindent\textit{Proof. }}

\newcommand{\qed}{\hspace{\stretch{1}$\square$}}

\newcommand{\sL}{ {\mathcal L}}

\newcommand{\sS}{ {\mathcal S}}

\newcommand{\medm}{ { \mu_{\m}} }
\newcommand{\medq}{ { \mu_{q}} }
\newcommand{\medi}{ { \mu_{i}} }
\newcommand{\Med}{ { \mu }}

\newcommand{\dist}{  \mbox{\textsf{d}} }
\newcommand{\polylog}{ {\mathrm{polylog} } }
\newcommand{\m}{1}
\newcommand{\threestate}{$\textrm{Undecided-State Dynamics}$}

\newcommand{\cdist}{\ensuremath{\mbox{\textsf{md}}(\bar \genc)}}
\newcommand{\cconf}[1]{\ensuremath{\langle c_1^{(#1)},\dots,c_k^{(#1)}, q^{(#1)}\rangle }}
\newcommand{\ccgen}{\ensuremath{\langle c_1,\dots,c_k, q\rangle }}
\newcommand{\genc}{\ensuremath{\mathbf{c}}}
\newcommand{\genC}{\ensuremath{\mathbf{C}}}

\newcommand{\End}{end}
\newcommand{\mdist}{\textnormal{\textsf{md}}}
\newcommand{\Q}{\mathcal Q}

\newcommand{\tm}[2]{\textrm{t}_\textrm{mix}^{#1}\left( #2 \right)}
\newcommand{\gossip}{$\mathcal{GOSSIP}$}
\newcommand{\local}{$\mathcal{LOCAL}$}

\newcommand{\ratio}{\ensuremath R}
\newcommand{\rratio}{\ensuremath \Lambda}

\newcommand{\proofof}[1]{\noindent\textit{Proof of~#1. }}

\newcommand{\ErdRen}{Erd\H{o}s-R\'{e}nyi }

\renewcommand{\leq}{\leqslant}
\renewcommand{\geq}{\geqslant}
\renewcommand{\le}{\leqslant}
\renewcommand{\ge}{\geqslant}
\begin{document}

\maketitle

\begin{abstract}
\input{./abstract.tex}

\end{abstract}
\thispagestyle{empty}
\newpage

\input{./trunk/intro.tex}

\input{./trunk/preliminaries.tex}

\input{./trunk/generalbounds.tex}

\input{./trunk/bigbang.tex}

\input{./trunk/afterbigbang.tex}

\input{./trunk/middleage.tex}

\input{./trunk/supremacy.tex}

\input{./trunk/expanders.tex}

\input{./trunk/conclusions.tex}

\newpage
\bibliographystyle{plain}
\bibliography{majority}

\newpage
\appendix

\input{./trunk/appendix.tex}

\end{document}

%% file: abstract.tex

We study   \emph{Plurality Consensus}   in the  \gossip\ 
\emph{Model} over  a network of $n$ anonymous agents. Each agent supports 
an initial opinion or    {\em color}.
We assume that at the onset, the 
number of agents supporting the \emph{plurality} color exceeds that
of the agents supporting any other color by a 
sufficiently-large    \emph{bias}, though the initial plurality itself might
be very far from absolute majority. The goal is to provide a 
protocol that, with high probability, brings the system into 
the   configuration in which all agents support the 
(initial) plurality color.
 
We   consider the   \emph{Undecided-State Dynamics}, a 
well-known  protocol which   uses 
just   one more state (the undecided one) than those   
necessary to store     colors.

We show that the speed of convergence of this protocol depends on the
initial color configuration as a whole, not  just  on the gap
between the plurality and the second largest color community. This dependence is
best captured by a novel notion we  introduce, namely,
 the \emph{monochromatic distance}  $\mdist(\bar \genc)$ which   measures
the distance of the initial color configuration $\bar \genc$ from the closest monochromatic one.  
 In the complete graph, we prove that, for a wide range of the  input 
parameters, this   dynamics      converges 
within $O(\mdist(\bar \genc) \log n)$   rounds. We   prove that this upper 
bound is almost tight in the strong sense: Starting from \emph{any} 
color configuration $\bar \genc$,  the convergence 
time is    $\Omega(\mdist(\bar \genc))$.

Finally, we   adapt   the \threestate \ to obtain  a fast, random walk-based
protocol for plurality consensus on \emph{regular expanders}. 
This protocol converges in $O(\mdist(\bar \genc) \,  \polylog (n))$
  rounds using   only $\polylog(n)$ 
local memory. A key-ingredient to achieve the  
above bounds is a new analysis of the maximum node congestion that results from 
performing $n$ parallel random walks on regular expanders. 

All our bounds hold with high probability.

\bigskip 
\noindent \textbf{ - Keywords.} Gossip Algorithms, 
Plurality Consensus, Markov Chains, Random Walks.

%% file: trunk/intro.tex
 \section{Introduction}

Reaching \emph{Plurality Consensus} is a fundamental task in 
distributed computing. Each agent of 
a distributed system initially supports a color, i.e. a number $i 
\in [k] = \{1, 2,\ldots , k\}$ (with $2 \leqslant k \leqslant n$). In the initial color 
configuration $\bar \genc = \langle \bar c_1, \ldots , \bar c_k \rangle$ (where 
$\bar c_i$ denotes the number of agents supporting color $i \in [k]$), 
there is an initial \emph{plurality} $\bar c_{\m}$ of agents supporting 
the \emph{plurality color} (wlog, we   assume that color communities 
are ordered, so that $\bar c_i \geqslant \bar c_{i+1}$ for any $i \leq k -1$). 
Initially, every agent only knows its own color; the goal 
is a distributed algorithm that, \emph{with high probability} (in short,
{\em w.h.p.})\footnote{As usual, we say that an event $\mathcal{E}_n$ holds 
w.h.p. if $\Prob{}{\mathcal{E}_n}\ge 1 - n^{-\Theta(1)}$.}, brings the 
system into the \emph{target} configuration, i.e., the monochromatic 
configuration in which all agents support the  initial plurality color. 
In the remainder, the subset of agents supporting color $i$ is called 
the \emph{$i$-color community}. 

\indent
This problem is also known 
as \emph{majority consensus} or
\emph{proportionate agreement} \cite{AAE07,AD12,PVV09}, but we prefer the term \emph{plurality} 
in this paper, in order to emphasize 
that the initial plurality $\bar c_{\m}$ might be far from the (absolute) majority: for instance, it could be   
some root of $n$. 
We study plurality consensus in the 
\emph{ \gossip\ model} \cite{CHKM12, DGHILSSST87,KDG03} in which each of $n$ 
agents of a communication network can, in every round, contact one (possibly random) 
neighbor to exchange information. 
Agents can be anonymous, i.e., they don't need to possess unique labels.
A major open question for the plurality consensus problem is whether a plurality 
protocol exists that converges in polylogarithmic time and uses only  polylogarithmic
local memory \cite{AAE07,AD12,PVV09}.

There is a strong interest for simple plurality 
protocols (called \emph{dynamics}) in which agents possess just a 
few more states than those   necessary to store the $k$ possible colors
\cite{AAE07,BCNPST13,CER14,DGMSS11,cardelli2012cell,PVV09}.
In this paper, we consider the \threestate\footnote{The Protocol has been initially 
``designed'' for the case $k=2$ and, thus, it has been named the \emph{Third-State} Dynamics.}, 
that  has been introduced in \cite{AAE07} and analyzed 
in \cite{AAE07,PVV09} only in the binary case (i.e. $k=2$).
The analysis of the multivalued case (i.e. $k >2$) has been proposed in \cite{AAE07,AD12,CER14,DGMSS11,Spir14} as an 
open problem.
The interest for this dynamics touches areas beyond the borders of computer science. 
It appears to play a major role in important biological processes
modelled as so-called
 chemical reaction networks \cite{cardelli2012cell,Doty14}. 

As discussed further in the introduction, in   previous work, the performance of 
this dynamics  on the complete graph has been  evaluated w.r.t. the 
following parameters: the number $n$ of nodes, the number $k$ of 
colors, and the initial \emph{bias} towards the plurality color, with the 
latter characterized in terms of a parameter that only depends on the 
relative magnitude\footnote{Typically, this relative magnitude is 
defined in terms of the absolute difference or the ratio.} of 
$\bar c_{\m}$ and $\bar c_{2}$.

\begin{figure}[!ht] 
\centering
\psfrag{bias}{\LARGE{$s$}}
\psfrag{av}{\LARGE{$n/k$}}
\psfrag{cc1}{\LARGE{Color configuration $\bar{\genc}$}}
\psfrag{cc2}{\LARGE{Color configuration $\bar{\genc'}$}}
\includegraphics[width=\linewidth,scale=0.75]{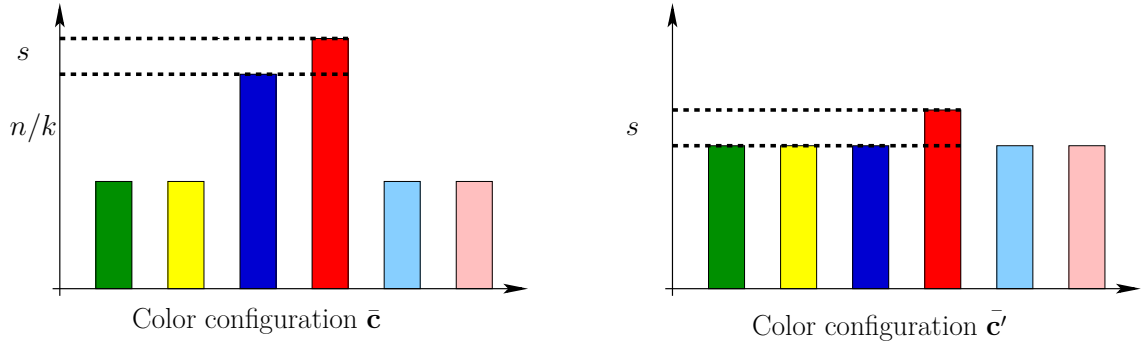}
\caption{Two different color configurations having the same bias $s= s(c_1,c_2)$ } \label{sod-fig:reg1full}
\end{figure}

However, when $k >2$, any such measure of the initial bias is not sensitive enough
to accurately capture the convergence time of a plurality protocol:
a \emph{global} measure is needed, i.e., one that reflects the 
whole initial color configuration. To better appreciate this issue, 
consider the two configurations $\bar \genc$ and $\bar \genc'$ in Fig. \ref{sod-fig:reg1full}. 
Whether the absolute difference or the relative ratio is used 
to measure the initial bias, the color configuration $\bar \genc'$ appears to be not ``worse'' than $\bar \genc$.
Still, computer simulations and intuitive arguments suggest that, under
any ``natural'' plurality protocol, 
the almost-uniform color distribution $\bar \genc'$ can result in
much larger  convergence times than the highly-concentrated color configuration $\bar \genc$. 

To the best of our knowledge, the impact of the whole  initial color configuration 
on the speed of convergence of plurality protocols has never been analyzed before.

\medskip
\noindent
\textbf{Our Contributions. }
We first 
introduce a suitable distance $\dist(\cdot,\cdot)$ (see Section \ref{se:prel}
for a formal definition) on the set $\sS$ of all color configurations. 
It naturally induces a function $\mdist(\cdot)$, called the \emph{monochromatic distance},
which 
equals the \emph{distance} between any configuration $\genc$ and the target configuration:

\[ 
	\mdist(\genc) \ = \ \sum_{i=1}^k \, \left(\frac{c_i}{c_{\m}}\right)^2    
\] 

We use $\mdist$ to characterize the bias of the initial configuration. In particular, note that 
$\mdist(\bar \genc) $ measures the extent to which $\bar \genc$ is ``uniform'': Indeed, the higher the
extent of the bias towards a small subset of the colors (including the plurality one), the smaller the value of $\mdist(\bar \genc)$.
As an example, in Fig. \ref{sod-fig:reg1full}, $\mdist(\bar \genc)$ can be substantially smaller than $\mdist(\bar \genc')$.
At the extremes, when there are only $O(1)$ color communities of size $\Theta(\bar c_{\m})$, we have
$\mdist(\bar \genc) = \Theta(1)$ while, when $\Theta(k)$ color communities have size $\Theta(n/k)$, we have
$\mdist(\bar \genc) = \Theta(k)$.

The simple strategy of the \threestate\ \cite{AAE07,PVV09} 
is to 
``add'' one extra   state to somewhat account for the ``previous'' opinion supported by an agent (see Section \ref{se:prel} 
and Table \ref{tab:The-update-rule} for a definition of this dynamics).
The convergence 
time of this dynamics has been analyzed on different distributed 
models, but only in the binary case \cite{AD12,AAE07,BD13,BTV09,DV12,PVV09}. In this restricted 
setting   the complex dependence of the dynamics' evolution on the
overall shape of the initial color configuration is not exhibited.

We analyse the \threestate\ using a technique that strongly departs
from past work and that allows us to address the plurality consensus problem in
the general setting. Our analysis 
achieves almost-tight bounds on convergence time. Formally,
let $k=k(n)$ be any function such that $k= O((n/\log n)^{1/3})$, and consider
any initial configuration $\bar \genc \in \sS$ 
such that $\bar c_{\m} \geqslant (1+ \alpha) \bar c_{2}$ where $\alpha>0$ is \emph{any arbitrarily-small} constant
(this is a weak-bias condition that 
ensures the convergence of the process towards the plurality color).
Then, the \threestate\ converges in
$O(\mdist(\bar \genc) \log n)$ rounds  w.h.p. 

This result is almost-tight in a
strong sense. In particular, we are able to prove that, for $k= O((n/\log n)^{1/6})$ and
\emph{for any} initial $k$-colors configurations $\bar \genc$, {\em the convergence time of the \threestate\ is
linear in the monochromatic distance $\mdist(\bar \genc)$  
w.h.p.}

The best previous results \cite{BCNPST13,KDG03} about plurality protocols 
will be compared to ours 
later in this introduction. We only emphasize that, when $k$ is some   root of $n$, 
our refined analysis implies that this dynamics 
is exponentially faster than the best protocol that uses polylogarithmic bounded memory
\cite{BCNPST13} on a large class of initial color configurations.
Moreover,  we observe that the \threestate\ uses exponentially-smaller message and memory size 
 w.r.t. the fastest (i.e.  polylogarithmic-time) gossip  protocol in \cite{KDG03}.

Our analysis   is rather general and it can be extended
to other interesting  topologies. As a case supporting this claim, we show how to 
adapt the Undecided-State Dynamics for the class of \emph{$d$-regular expanders} 
\cite{HLW06}, for any degree $d \geqslant 1$. 
Efficient dynamics for this class of graphs have only been  analyzed  for the binary case
\cite{CER14,MNT12}.

In this variant of the \threestate, the task of selecting random neighbors is simulated by performing $n$ independent 
random-walks of suitable length. Thanks to the well-known
rapidly-mixing properties
of $d$-regular expanders \cite{HLW06,lpwAMS08}, we can prove that
the new protocol converges in time $O(\mdist(\bar \genc) \polylog ( n))$, w.h.p. 

The major technical hurdle here is proving that this variant of the 
protocol still requires $\polylog(n)$ local memory. To this aim,
we prove that the \emph{node congestion} is at most $\polylog(n)$.
The analysis of the process that results from running parallel random walks over 
a graph has been the subject of extensive research in the past \cite{AAKKLT08,FKP11,HPPRS12,P00,SMP12}.
However, to the best of our knowledge,
none has addressed the issues we consider here. In particular, the analysis of node congestion 
is far from trivial and of independent interest, since 
efficient protocols for several important tasks in the \gossip\ model (such as \emph{node-sampling} \cite{SMP12},
\emph{network-discovery} problems
\cite{HPPRS12}, and \emph{averaging} problems \cite{BGPS06}) rely on the use of parallel random walks.

\medskip
\noindent
\textbf{Motivations and comparison to previous works. }
Plurality consensus (a.k.a. majority consensus or proportionate agreement) 
is a fundamental problem 
arising in several areas such as \emph{distributed computing} \cite{AAE07,DGMSS11,Pe02}, \emph{communication networks} \cite{PVV09},
\emph{social networks} \cite{CDGNS13,MS10,MNT12} and \emph{biology} \cite{cardelli2012cell}. 

\smallskip
Applications include fault-tolerance in parallel computing and in distributed database management where 
data redundancy or replication and majority-rules are used to manage the presence of unknown
faulty processors \cite{DGMSS11,Pe02}. 
Another application comes from the
task of distributed item ranking, in particular when every node initially 
ranks some item and the goal is to 
agree on the rank of the item based on the initial plurality opinion \cite{PVV09}. 
Further areas of interest of the multi-valued case include distributed 
cooperative decision-making and control in environmental monitoring, 
surveillance and security \cite{RM08}. Finally, converging to the plurality color 
among a (large) set of initial 
node colors has been recently used as a basic building block for \emph{community detection} in dynamic social networks
\cite{CDGNS13}. 
We remark that, in all such applications, the data domain 
can span a relatively-large range of values, hence
the importance of this problem for large values of $k$.

\smallskip
Interestingly enough, only the binary case is essentially settled,
even for complete graphs. In the synchronous model, a simple gossip protocol 
for computing the median can be used to solve the majority consensus problem
in the binary case,
with constant memory and message size \cite{DGMSS11}. The proposed protocol converges in $O(\log n)$ time rounds
if the initial difference bias $s = \bar c_{\m} -\bar c_{2}$ is $ \Omega(\sqrt{n \log n})$.

More recently,
in \cite{CER14}, the authors provide a rigorous analysis of a simple 2-voting dynamics for the binary case
on any (possibly random) regular graph: in the latter case, they provide optimal bounds
on the convergence time as a function of the second-largest eigenvalue of the graph. 

For the multivalued case, in \cite{BCNPST13} the authors analyze a gossip 
protocol, called \emph{$3$-Majority Dynamics}, where at every round, each agent applies a 
simple majority rule over the colors of three randomly-sampled neighbors. 
When the initial difference bias is
$s = \Omega(\sqrt{kn \log n})$, 
the $3$-Majority Dynamics converges in $\Theta(\min\{k, n^{1/3}\} \log n)$ rounds using $\Theta( \log k )$ memory and message size.
 
Convergence times of the $3$-Majority Dynamics become polylogarithmic only if $\bar c_{\m} \geqslant n/\polylog(n)$,
thus they are not polylogarithmic 
whenever $k = \omega(\polylog(n))$ and $\bar c_{\m} = o(n/ \polylog(n))$. This is the parameter range where
our analysis of the \threestate\ leads to an exponential speed up w.r.t. the convergence time
of the $3$-Majority Dynamics. For example, consider an initial  ``oligarchic''  scenario where  $k = n^{1/4}$ and 
  a subset $\sL \subseteq 
[k]$ exists such that $|\sL| = \polylog(n)$,  for any $i \in \sL$,  $\bar c_i \sim n/\sqrt{k}$, and,
  for any $i \in [k] \setminus \sL$,   $\bar c_i \sim n/k$. Clearly, 
$1,2 \in \sL$ and the resulting monochromatic distance is
$\mdist(\bar \genc) = \polylog (n)$. Assuming  
$\bar c_{\m} \geqslant (1 + \alpha) \bar c_{2}$ for some $\alpha >0$    our upper bound implies that, 
starting from any such configuration, the \threestate\   converges in 
polylogarithmic time, whereas the 3-Majority Dynamics converges in $\Theta(k\log n)$ time  \cite{BCNPST13}.

In \cite{KDG03}, the authors provide
a gossip protocol to compute aggregate functions, which
can be used to solve  plurality consensus in $\polylog(n)$ time
starting from any positive bias, 
but it requires exponentially larger memory and message size (namely $\Theta(k\log n)$). 
The \threestate\ has been introduced and analyzed 
in \cite{AAE07} for the binary case in the 
population protocol model (where only one edge is active during a   round). 
They prove that this dynamics has ``parallel'' convergence time 
  $O(\log n)$
whenever the bias  $\Omega\left(\sqrt{n \log n}\right)$. 
In \cite{BD13,BTV09,DV12,PVV09,Spir14}, the same dynamics for the binary case has been analyzed 
in   different distributed models.
Last but not least, interest for this dynamics was stimulated 
by recent findings in biology: notably, as shown in \cite{cardelli2012cell}, 
the structure and dynamics of the ``approximate majority''
protocol (as it is called there and in \cite{AAE07}) is to a great extent similar  
to a   mechanism that is collectively implemented in the network that regulates the mitotic
entry of the cell cycle in eukaryotes. 

We   mention that similar majority-consensus problems have been 
studied (for example  in \cite{AD12,MNT14}) in the  \emph{\local\ (communication) model} 
\cite{FKP11,P00} where, however, node congestion and   memory size are 
linear in the node degree of the network.

%% file: trunk/preliminaries.tex
\section{Preliminaries}\label{se:prel}

Let us consider a complete graph  of  $n$ anonymous nodes (\emph{agents}): each of them  is initially 
colored with one out of $k$ possible colors, where $k = k(n) \in [n]$. It is assumed that
there is an initial \emph{plurality} $c_{\m} > n/k$ of agents  colored with the \emph{plurality color} $\m$.  
A synchronous   protocol for the \emph{plurality problem} is a finite  set  of local rules (applied by 
every agent) that bring  the system
into  the \emph{target}  configuration where all agents are colored by $\m$.

\medskip
\noindent
\textbf{The \threestate .}
We analyze the synchronous version of the  \threestate\  introduced in~\cite{AAE07} and \cite{PVV09}. 
Differently from other ones (e.g., the majority dynamics considered in \cite{BCNPST13}), in this protocol,
after the first round, agents can also enter an \emph{undecided} state $q$, to which no color is associated. 
Accordingly, at each round $t$, the global state
of the system can be represented  by a \emph{color configuration} $\genc^{(t)}=\cconf{t}$ where $c_{i}^{(t)}$ is the number of $i$-colored agents, and $q^{(t)}$ is the number of undecided agents. 
In the sequel, wlog, we will assume that $c_i \geq c_{i+1}$ for any $i \leq k-1$.

The   \threestate\    works as follows.
According to the (uniform) gossip model,
at  every round $t \geqslant 0$, each agent $u$ chooses     a neighbor $v$ uniformly at random and 
decides to get   a new color/state according to the rules in Table \ref{tab:The-update-rule},

\begin{table}[h!]
\centering
\begin{tabular}{|c||c|c|c|}
\hline 
$u\big\backslash v$ & undecided & color $i$ & color $j$\tabularnewline
\hline 
\hline 
undecided & undecided & $i$ & $j$\tabularnewline
\hline 
color $i$ & $i$ & $i$ & undecided\tabularnewline
\hline 
color $j$ & $j$ & undecided & $j$\tabularnewline
\hline 
\end{tabular}
\caption{\label{tab:The-update-rule}The update rule of the \threestate\ 
where $i,j\in\left[k\right]$ and $i\neq j$.}
\end{table}

The dynamic process that results from running the \threestate\ on the complete
graph can be represented by a finite-state Markov chain defined 
over the space of all color configurations. In the next subsection, 
we formally define     this Markov 
chain and our concept of global bias.

\subsection{Basic definitions and global bias}
We next provide the basic notation and conventions
adopted in this work, give some key definitions
and discuss some preliminary facts that will be useful in the
remainder.

\smallskip
\noindent{\bf Basic notation.}
Considered any time $t$, the state of the process (i.e.  
the  Markov chain) is completely   characterized 
by the corresponding color configuration, namely by  

\[ \genc^{(t)}=\langle c_{1}^{(t)},c_{2}^{(t)}, \ldots,c_{k}^{(t)},q^{(t)}\rangle  \]
The set of all possible color configurations will be denoted by $\sS$.
In the initial state we always have $q^{(0)} = 0$.

For any time $t \geqslant 0$, 
  the execution of one round of the    dynamics rule  (uniquely) determines the probability distribution of 
the (vectorial) random variable representing the   random state at time $t$:

\[ \genC^{(t)}=\langle C_{1}^{(t)},C_{2}^{(t)}, \ldots,C_{k}^{(t)},Q^{(t)}\rangle  \]
Notice that we omit in the notation the dependence of the random state on the initial color configuration.
This random process is clearly a   finite-state  Markov chain.

In general, 
lower-case letters will be used to denote functions of
the observed  color configuration at any specified time.
Upper-case letters instead will denote \emph{random variables}.

Thus, $Q^{(t)}$ and $C_i^{\left( t  \right)}$  denote
the r.v.s counting the  number of nodes that, respectively, are undecided   and that  have color $i$ at time $t$.

If we condition the system to be in a fixed state $\genc$ at a generic round,
  the random sizes of the $i$-color communities and that of the undecided community at the next round will 
  be denoted as  $C_{i}'$ and $Q'$, respectively.
  
For brevity' sake, we define
	\begin{align*}
		\medi :=  \expec{C_{i}'}{\genc} \,\,\, (i \in [ k ]),
			\quad
		\Med  := \expec{\sum_iC_i'}{\genc} 
			\quad \mbox{ and } 
			\quad
		\medq :=  \expec{Q'}{\genc} 
	\end{align*}

Finally, we often write $\mathbf{P}\left(A\right)$ for
$\mathbf{P}\left(A|B\right)$ when the conditioning event $B$ is
clear  from context. 
 
\smallskip
\noindent{\bf Global bias.} 
Our analysis will  highlight a fundamental  dependence of convergence properties of the \threestate\ 
on a particular measure of the initial global bias. 
To mathematically characterize  this we next  introduce the following notion of  distance
between \emph{equivalent} color configurations.

Given any color configuration $\genc\langle c_1,c_2,\ldots,c_k, q\rangle$,  
 consider  the following ratio

\[    R(\genc) \ = \  \sum_{i=1}^{k}  \frac {c_i}{c_{\m}}    \] 
 This  allows us to define  an equivalence relation $\equiv$ in the space $\sS$

\[ \genc\,  \equiv \, \genc'  \  \ \mbox{ iff } \  \  R(\genc) = R(\genc') \]
 and the following  function   over pairs of  equivalence classes  (with an abuse of notation, for any
  color configuration $\genc$, we
  will denote its equivalence  class as $\genc$ as well)

\begin{equation*} 
	d\left(\genc,\genc'\right) \ =\ \sum_{i}\left(\frac{c_{i}}{{c}_{\m}}-
	\frac{c_{i}'}{{c}_{\m}'}\right)^{2}
\end{equation*}

It is easy to verify that the function $d(\cdot,\cdot)$ is   a distance over the quotient space of $\sS$.
Let us now consider the equivalence  class  $\mathcal M$ of the ($k$) possible \emph{monochromatic} color configurations
and   recall the definition of \emph{monochromatic distance} (given in the introduction), 
\[ \mdist(\genc) \ = \ \sum_{i=1}^k \, \left(\frac{c_i}{c_{\m}}\right)^2      \] 
   Then, 
  we immediatly have 
\[ 
\mdist\left(\genc\right) \ = \ d(\genc, \mathcal M) + 1
\]

The simple considerations above entail that $\mdist$ defines a notion of distance from
the monochromatic configuration that corresponds to the initial plurality. Consistently, it is
straightforward to see that $\mdist$ is maximized by ``uniform'' configurations, i.e., configurations 
$\genc$ such that $c_{\m}\approx n/k$.
For every $\genc$, it holds that 
\begin{equation}\label{fa:ranger}
1 \ \leq \ R (\genc), \mdist(\genc) \  \leq  \ k 
\end{equation}

\noindent
Finally, let us define the following ratio
\[   
	\rratio(\genc) \  :=  \ \frac{ R(\genc )^2 }{ \mdist(\genc) }
\] 
From the definitions of $R(\genc)$ and $\mdist(\genc) $ and 
		from a simple application of the  Cauchy-Schwartz 
		inequality to $R(\genc)$, we get 
\begin{equation}\label{le:ratio1} 
	\rratio(\genc) \  \le \  k 
\end{equation} 
for every configuration $\genc$.

\section{Analysis of the \threestate} \label{ssec:overvfull}
The presence of an extra, undecided state makes the analysis hard and interesting. The evolution of the system
does quantitavely depend on the initial configuration but, when the initial  bias is high enough, 
w.h.p. the evolution of the system follows a ``typical'' pattern, characterized by some 
consecutive phases  with pretty different regimes. 
In fact, the typical evolution
of the system is relatively simple to describe.

\begin{figure}[h!]
	\centering
	\includegraphics[width=\linewidth]{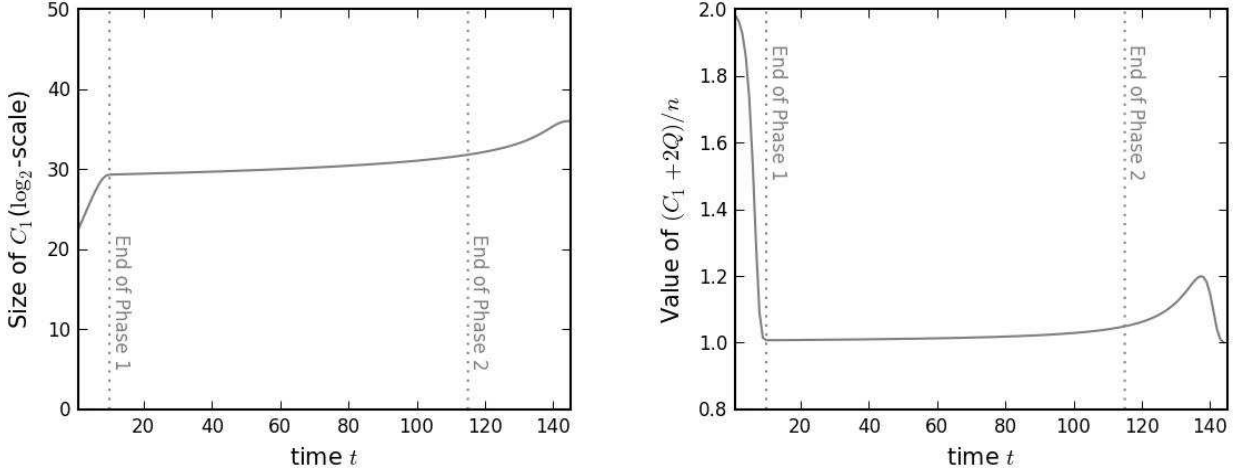}
	\caption{Typical evolution of the \threestate{} after the first round, for $n=7\cdot 10^{10}$ nodes and $k=(\frac{n}{\log n})^{\frac{1}{4}}$ colors, with $c_{\m}^{(0)}=2\frac{n}{k}$ and $c_i^{(0)}=\frac{n}{k}\left( 1-\frac{2}{k} \right)$ for every $i\neq \m$.}
	\label{fig:evol}
\end{figure}

On the other hand, understanding the reasons that determine this pattern of
behaviour requires an analysis that is far from trivial. In the next 
Subsection, we provide a high-level description of the typical
evolution of the process when the initial bias is high enough.

\subsection{The process in a nutshell}\label{subse:nutsfull}
The typical behaviour of the \threestate \ is exemplified in Fig. \ref{fig:evol}.
The typical process evolution appears to unfold across the first round and, then,  three different phases.  

\smallskip
\noindent
\textbf{First Round: \textit{Rise of the undecided.}}
After the  first round, we see
  dramatic changes in the system: i) in 
general, a drastic drop in the sizes  of all color communities occurs,
with color communities whose initial size  is $o(\sqrt{n})$ simply 
disappearing w.h.p.; ii) a large fraction (possibly 
the vast majority) of undecided nodes emerges; iii) under 
reasonable bias assumptions, the initial plurality color does not
change w.h.p., though its size  drastically 
drops in absolute terms.  

\smallskip
\noindent
\textbf{First phase: \textit{Age of the undecided.}}
This phase starts right after round 1. The duration of this phase
actually depends on a non obvious function of the initial 
color configuration (and not just the magnitude of the initial bias)
and it can range from $O(1)$ to $O(\log n)$ rounds. 
During this phase, w.h.p., the sizes of the various
color communities grow (almost) exponentially fast. At the same time,
the relative ratios between the   
  plurality and    the other-community sizes are
approximately preserved w.h.p.. This phase ends as the following events w.h.p. 
occur: i) $C_{\m}$ becomes 
$\Omega(n/\mdist(\bar \genc))$ and ii)  $Q$  drops
to a value slightly above $n/2$. 

\smallskip
\noindent
\textbf{Second phase: \textit{Plateau} or \textit{Age of stability.}}
  During this phase, $C_{\m}$ increases roughly at a rate $1 + 
\Theta(1/\mdist(\bar \genc))$, while the \emph{ratios}  $C_{\m}/C_i$ are preserved. 
Under the typical configurations in which the system is
at the end of the  previous phase,  this  phase is characterized
by a relatively long ``plateau'' lasting
$\Omega(\mdist(\bar \genc))$ rounds w.h.p., during which plurality remains
stuck around a cardinality value $\Theta(n/\mdist(\bar \genc))$.

\smallskip
\noindent
\textbf{Third phase: \textit{From plurality to totality.}}
Though sub-exponential, the expected minimal growth rate of $C_{\m}$ during the previous phase allows the
plurality to increase, so that within $O\left(\mdist(\bar \genc) \log n\right)$ rounds,
 $\mdist(\genC)= 1+o(1)$.  It is possible to see that, from this point onward,
$\sum_{i\neq \m}C_{i}+Q$  decreases exponentially fast in expectation, till the end of the process.

%% file: trunk/generalbounds.tex
\subsection{General bounds}\label{se:genbounds}
Before describing the process' phases, we here  provide some crucial properties
that hold along the entire process.
  
If $\genc = \ccgen$ is the current color configuration (i.e. state)
of the Markov chain, then we can easily derive the ``expectation'' of the  
next color configuration
\begin{align}
	\medi & =  \expec{C_{i}'}{\bar{\genc}} = c_{i} \cdot \frac{c_{i}+2q}{n} \label{eq:average_of_c} \quad (i \in \left[ k \right])\\
	\medq & = \expec{Q'}{\bar{\genc}} = \frac{q^{2}+\sum_{i\neq j}c_{i}\cdot c_{j}}{n}=\frac{q^{2}+\left(n-q\right)^{2}-\sum_{i} c_{i}^{2}}{n}\label{eq:average_of_q}
\end{align}

\noindent
From   (\ref{eq:average_of_c}), we can see the crucial role of the quantity  $\frac{c_{i}+2q}{n} $: it 
in fact represents the expected \emph{growth rate} of every color community. 
A major novelty of our contribution is 
 
\smallskip
\begin{verse}
\emph{the discovery of a clean 
mathematical connection between   the expected growth rate of   plurality   and   
the monochromatic distance of the current configuration.}
\end{verse}
  
 The  following lemma in fact formalizes such a connection by means of    $\ratio(\genc)$ and it   
     plays a  key role  in our  analysis of the entire process evolution. As will see in Lemma \ref{lem:R_first_step},
     $\ratio(\genc)$ and $\mdist(\genc)$ are in fact strongly related.

	\begin{lemma}[Plurality Drift]
		\label{lem:maj_drift}
		Assume that, at some round,  the system is in a   color configuration 
		$\genc$  such that 
		$c_{\m} \geq \left( 1+\alpha \right)c_i$ for any $i \neq \m$ and for some constant $\alpha>0$.
		Then, at the next round, it holds that 
		\begin{equation*}
			\expec{\frac{ C_{\m}' + 2Q' }{n}}{\genc}  \ \geq \	1+\Gamma(\genc)
		\end{equation*}
		where \[\Gamma(\genc)=\left( 1-\frac{ c_{\m} +2q }{n} \right)^{2}+2\left( 1-\gamma \right)\left( \ratio(\genc) -1 
		\right)\left(\frac{ 
		c_{\m} }{n}\right)^{2}  \ \mbox{ with }   \  \gamma=\left( 1+\alpha \right)^{-1}  \]
	\end{lemma}
		
		\proof
			Let $\beta=(1-\gamma)$. By using the hypothesis 
			\[  
				\frac{c_i}{c_{\m}}\leq\frac{1}{\left( 1+\alpha \right)} 
			\] 
			we get
			\[
				\mdist(\genc)= \sum_i \frac{ c_i^2}{ c_{\m}^2} \leq 1 + \frac{1}{\left( 1+\alpha \right)}\sum_{i\neq 1} \frac{c_i }{c_{\m} } 
					= \gamma \ratio(\genc)+\beta
			\]
			Moreover, we can write $q$ as  $q=  n-\ratio(\genc)c_{\m}$. Thanks to the above equations and    
			 (\ref{eq:average_of_c}) and (\ref{eq:average_of_q}), by simple manipulations, we get 
			\begin{align*}
				&\expec{ \frac{ C_{\m}'+2Q'}{n}}{\genc} = c_{\m}\cdot \frac{c_{\m}+2q}{n^2}+2\frac{q^2+
				\left( n-q \right)^2-\sum_i (c_i)^{2}}{n^2} \\
				&= c_{\m}\cdot \frac{c_{\m}+2q}{n^2}+2\frac{q^2+\left( \ratio(\genc)^{2}-\mdist(\genc) \right)\cdot (c_{\m})^{2}}{n^2} \\
				 &\geq c_{\m}\cdot \frac{c_{\m}+2q}{n^2}+2\frac{q^2+\left( \ratio(\genc)^{2}-\gamma \ratio(\genc)-\beta\right)
				\cdot (c_{\m})^{2}}{n^2}= \\
				& =   \frac{c_{\m}^2+2\left( n-\ratio(\genc)c_{\m} \right)+2\ratio(\genc)^{2}(c_{\m})^{2}+
				2\left( n-\ratio(\genc)c_{\m} \right)^2-2\gamma \ratio(\genc)(c_{\m})^{2}-2\beta(c_{\m})^{2}}{n^2}\\
				& = 1+\left( 1 - \frac{c_{\m}+2q}n\right)^2+2 \left( 1-\gamma \right) \left(\ratio(\genc) -1\right)\frac{c_{\m}^2} {n^2}
			\end{align*}
		\qed

Another   useful property   that is often used in  our analysis    is the fact that some crucial r.v.s
are essentially monotone along the entire process. In the next lemma, we prove this monotonicity
for the r.v.s  
$\ratio(\genC')$ and the ratios $C_i'/C_{\m}'$ (for $i \neq \m$).

	\begin{lemma}[Monotonicity]\label{lem:R_any_step}
		Assume that, at some round,  the system is in a   color configuration 
		$\genc$  such that, for some constant $\alpha>0$ and a large enough constant $\lambda>0$ it holds 
		\[ 
			c_{\m}\geq\left(1+\alpha\right)c_{i}
			\, \mbox{ for   any }  i \neq \m \ \mbox{ and } 
			\  \medm \geq \lambda\log n 
		\]
		Then, at the next round, w.h.p.
		it holds that: 
		\begin{align}
			&\ratio(\genC') < \ratio(\genc) \cdot\left(1+{ O\left(\sqrt{\frac{\log n}{\medm}}\right) }\right)
			\label{eq:monotone_r}\\
			&C_{\m}'\geq\left(1+\alpha\right)\cdot C_i'\cdot\left(1- O\left(\sqrt{\frac{\log n}{\medm}}\right) \right)
			\label{eq:monotone_bias}
		\end{align}
	\end{lemma}
		\proof
			As for Claim \eqref{eq:monotone_r}, since $\ratio(\genC')=\frac{\sum_iC_i'}{C_{\m}'}$, it suffices to   
			bound,   respectively,       $C_{\m}'$  and $\sum_iC_i'$.
			By applying the Chernoff bounds  \eqref{eq:CB_lower_multip} and \eqref{eq:CB_upper_multip}
			and by using  the hypothesis $\Med \geq \medm \geq \lambda\log n$
			we get
			\begin{align}
				&\probc{C_{\m}'\leq \medm
					\cdot\left(1-\sqrt{\frac{2a\cdot\log n}{\medm}}\right)}{\genc}
					\leq\frac{1}{n^{a}}
				\label{eq:lower_on_c_1}\\
				&\probc{C_{\m}'\geq\medm
					\cdot\left(1+\sqrt{\frac{3a\log n}{\medm}} \right)}{\genc}
					\leq \frac{1}{n^{a}}
				\label{eq:upper_on_c_1}\\
				&\probc{\sum_iC_i'\geq\Med\cdot\left(1+\sqrt{\frac{3a\log n}{\Med}}\right)}{\genc}
					\leq \frac{1}{n^{a}}
				\label{eq:upper_on_sum_c_i}
			\end{align}
			for any constant $a \in \left( 0, \frac \lambda 3 \right)$.

			\noindent
			Let $A$ be the event in (\ref{eq:lower_on_c_1}),
			let $B$ be the event in  (\ref{eq:upper_on_sum_c_i}) and let $A^c$ and $B^c$ be their complimentary events, respectively. Observe that, from Lemma \ref{lem:wop_intersection}, it holds 			
			$\mathbf{P}\left(A^{c}\cap B^{c}\right)\geq1-\frac{2}{n^{a}}$.
			Moreover, by using that 
			\[
				\frac{1+\sqrt{\frac{3a\log n}{\Med}}}{1-\sqrt{\frac{2a\log n}{\medm}}}
				\leq 
				\frac{1+\sqrt{\frac{3a\log n}{\lambda \log n}}}{1-\sqrt{\frac{2a\log n}{\lambda \log n}}}
				\leq 
				{1+\sqrt{\frac{ba\log n}{\lambda \log n}}}
				\quad
				\mbox{ where }
				\quad
				b = \left( \frac{\sqrt{3}-\sqrt{2}}{1-\frac{3\sqrt{2}a}{\lambda}} \right)^2
			\]
			Using the latter two facts, we have that
			\begin{align*}
			& \probc{\ratio(\genC') = \frac{\sum_iC_i'}{C_{\m}'} < \frac{\sum_ic_{i}}{c_{\m}} 
				\cdot \left(1+\sqrt{\frac{ba\log n}{\Med}}\right)}{\genc} \geq \nonumber\\
			& \geq \probc{ \frac{\sum_iC_i'}{C_{\m}'} < \frac{\sum_ic_{i} 
				\cdot \left( c_{i} + q \right)}{c_{\m}\cdot \left( c_{\m}+ q \right)} 
				\cdot \left(1+\sqrt{\frac{ba\log n}{\Med}}\right)}{\genc}=\\ 
			& = \probc{\frac{\sum_iC_i'}{C_{\m}'} < \frac{\Med}{\medm}
				\cdot\left(1+\sqrt{\frac{ba\log n}{\Med}}\right)}{\genc} \geq\\ 
			& \geq \probc{\frac{\sum_iC_i'}{C_{\m}'} < 
				\frac{\Med\cdot\left(1+\sqrt{\frac{3a\log n}{\Med}}\right)}
				{\medm\cdot\left(1-\sqrt{\frac{2a\log n}{\medm}}\right)} }{\genc} 
				\geq \mathbf{P}\left(A^{c}\cap B^{c}\right)\geq1-\frac{2}{n^{a}}
			\end{align*}

			\noindent
			As for Claim \eqref{eq:monotone_bias}, the hypothesis $c_{\m}\geq\left(1+\alpha\right)c_{i}$ 
			clearly implies $\medm\geq\left(1+\alpha\right)\cdot \medi$. 
			Thus, by  \eqref{eq:lower_on_c_1}  we get
			\begin{align}
				\probc{C_{\m}'\leq\left(1+\alpha\right)\cdot \medi\cdot\left(1-\sqrt{\frac{2a\log n}{\medm}}\right)}{\genc}
				\leq \probc{C_{\m}'\leq \medm\cdot\left(1-\sqrt{\frac{2a\log n}{\medm}}\right)}{\genc}\leq
				\frac{1}{n^{a}}
				\label{eq:bias_conservation}
			\end{align}

			\noindent
			We now consider  two cases. 
			If $\medi < {\medm}/(6\left( 1+\alpha \right))$
			then, by Chernoff bound \eqref{eq:CB_absolute} (choosing $\delta={\medm}/{\left( 1+\alpha \right)}$), 
			with probability $1-n^{-\frac{\lambda}{1+\alpha}}$ it holds that 
			$C_i'\leq {\medm}/(\left( 1+\alpha \right)) $.
			Together with \eqref{eq:lower_on_c_1}, this implies that w.h.p. 
			\[
				C_{\m}' > \medm\cdot\left(1-\sqrt{\frac{2a\log n}{\medm}}\right)
					>\left( 1+\alpha \right)C_i'\cdot\left(1-\sqrt{\frac{2a\log n}{\medm}}\right)
			\]
			On the other hand, if $\medi\geq {\medm}/(6\left( 1+\alpha \right))$ 
			then, from the Chernoff bound \eqref{eq:CB_lower_multip} we get that 
			\begin{equation}
				\probc{C_i'\geq\medi\cdot\left(1+\sqrt{\frac{3a\log n}{\medi}}\right)}{\genc} 
				\leq \probc{C_i'\geq\medi\cdot\left(1+\sqrt{\frac{3a\log n}{\medm/6(1+\alpha)}}\right)}{\genc}
				\leq \frac 1{n^{a}}
				\label{eq:bias_conservation2}
			\end{equation}
			for any $a\in \left( 0,\frac{\lambda}{18\left( 1+\alpha \right)} \right)$.
			Thus, by using \eqref{eq:bias_conservation}, \eqref{eq:bias_conservation2} and Lemma \ref{lem:inequality} we get that w.h.p.
			\[
				C_{\m}'\geq\left(1+\alpha\right)\cdot C_i'\cdot\left(1-O\left(\sqrt{\frac{\log n}{\medm}}\right)\right)
			\]
		\qed

%% file: trunk/bigbang.tex
\subsection{First Round: \textit{Rise of the undecided}}
After the first round, a  strong decrease
of  the color communities happens, while  the undecided community   get to a large majority of the agent.
 
The next lemmas provide some formal statements about 
 this   behaviour which represent the  key start-up of  the process
(and its analysis). 

 We will implicitly assume that   the process starts in a fixed initial 
color configuration
 \[ \bar{\genc} =\langle \bar c_{1} ,\bar c_{2} , \ldots,\bar c_{k} \rangle \]
So, in the next lemmas, events and related probabilities  are conditioned on some fixed 
$\bar{\genc} $.

We observe that when $k$ is large, i.e. when $k = \omega\left( n^{b} \right)$ for some  $b\in (\frac 12, 1]$,
 if the process starts from    ``almost-uniform''    color configurations
	then, after the first round, even the plurality may disappear (w.h.p.): indeed, if we consider any 
	$\bar \genc $ such that $\bar c_{\m} = O \left( \frac nk \right)$, then a simple application of the 
	Markov inequality implies that $C_{\m} ' = 0$ w.h.p.
We will  thus focus on   ranges  of $k$ such that $k <  \sqrt{n/\log n}$.

	\begin{lemma}
		\label{lem:first_step_bounds}
			Let $k = o\left(\sqrt{n/\log n}\right)$. Given any initial color configuration $\bar \genc $, after the first round w.h.p. it holds:
		\[ 
		\frac{1}{2} \,  \frac{n}{ \ratio(\bar \genc)^2} \ \leqslant \  C_{\m}'  \leqslant \  2 \,  
		\frac{n}{ \ratio(\bar \genc)^2}\label{eq:fsc1} \ \mbox{ and } \ 
		{n}\left(1 - \frac{2} {  \rratio(\bar \genc)} \right) \ \leqslant \ Q' 
		 \ \leqslant \  {n}\left(1 - \frac{1} { 2 \rratio(\bar \genc)} \right)   
		\]
	\end{lemma}
		\proof
			From (\ref{eq:average_of_c}) and recalling that in the initial configuration $q  = 0$, we get 
			$$
				\medm = \frac{(\bar c_{\m} )^2}{n} = \frac{n}{\ratio(\genc)^2}
			$$
			Similarly, from (\ref{eq:average_of_q}) we get 
			\begin{eqnarray*}
				&&\medq = \frac{n^2 - \sum_i\left(\bar c_{i} \right)^2}{n} 
			= \frac{n^2 - \cdist \cdot \left(\bar c_{\m} \right)^2}{n} = n\left(1 - \frac{1}{\rratio(\bar{\genc})} \right)
			\end{eqnarray*}
			where the second equality follows from the definition of $\cdist$, while the third one
			from the definition of $\ratio(\bar \genc)$ and from simple manipulations.
			Since we assumed $k \leq o\left(\sqrt{n/\log n}\right)$ then we have that 
			\begin{eqnarray*}
				&&\medq = \frac{n}{\ratio(\bar \genc)^2}\ge \frac{n}{k^2} = \omega(\log n)
			\end{eqnarray*}
			The above inequality 
			allows us to apply the Chernoff bound  and prove the first  claim (i.e. that on $C_{\m}'$).
			 
			\noindent
			Similarly, 
			from \eqref{le:ratio1}, it holds 
			\[ 
				\frac{ n }{  \rratio(\bar \genc)}\ge\frac{n}{k} 
			\]  
			This allows us to apply the additive version of the Chernoff's bound and   prove the second claim (i.e that on   $Q'$).		
		\qed

\smallskip
The next lemma relates $\ratio(\genc)$ to $\mdist(\bar \genc)$
after the first round.
	\begin{lemma} 
		\label{lem:R_first_step}
		Let $k = o\left( \sqrt{n/\log n} \right)$. Given any initial color configuration $\bar \genc $, 
		after the first round w.h.p. it holds 
		\[   \ratio(\genC^{(1)}) \ \leq \  \mdist(\bar \genc)\cdot\left(1+ o(1)\right) \]
	\end{lemma}
		\proof
		By definition of plurality color, it holds that 
		$c_{\m} > n/k$. Therefore, by the hypothesis on $k$ and  (\ref{eq:average_of_c}), we get
		$\medm= \omega(\log n)$ and then, by using the Chernoff bounds of Lemma \ref{lem:cbmult}, we can get  concentration bounds
		on both    the numerator and the denominator of $R(\genC^{(1)})$ (as we did in 
			the proof of Lemma \ref{lem:R_any_step}).   Formally, we have that w.h.p.  
			\begin{align*}
				R(\genC^{(1)}) = \frac{\sum_i C_i^{(1)}}{C_{\m}^{(1)}} 
					\leq \frac{\Med}{\medm} 
					\cdot\left(1+ o(1) \right)
			\end{align*}
			Observe that, since in the initial color configuration $q =0$, it holds 
			\[
				\frac{\Med}
					{\medm}
				=\frac{\sum_{i}\left(\bar c_{i} \right)^{2}}
					{\left(\bar c_{\m} \right)^{2}}
			\]
			It follows that w.h.p.
			\begin{align*}
				R(\genC^{(1)}) \leq \frac{\Med}{\medm} 
					\cdot\left(1+ o(1) \right) 
				= \frac{\sum_{i}\left(\bar c_{i} \right)^{2}}
					{\left(\bar c_{\m} \right)^{2}}\cdot\left(1+o(1)\right) = \cdist\cdot\left(1+o(1)\right)
			\end{align*}
		\qed

%% file: trunk/afterbigbang.tex
\subsection{ First phase:  \textit{Age of the undecided}}

In this  phase, the undecided community rapidly decreases to a value close to $n/2$
while   the plurality reaches a size close to $n/(2\cdist)$. When this happens,
 the    ratios $C_i/C_{\m}$ and $ R{(\genc)} $ will essentially keep their initial values and 
 the $Q$  will decrease to a value very close to $n/2$. 
The length of this phase is at most logarithmic. 

The next lemma formalizes the aspects of this phase that will be used to get the upper bound on the convergence time of the process.

	\begin{lemma} \label{lem:ub_phase2} 
	Let  $k = o\left( \sqrt{{n}/{\log^2 n}} \right)$ and let $\epsilon$ be any constant in $(0,\frac 12)$. 
		Let  $\bar \genc$ be any initial configuration such that, 
		for any $j \neq \m$ and for some arbitrarily small constant $\alpha>0$,  
		  $c_{\m}  \geq  \left( 1+\alpha \right) \cdot c_{j}$.
		Then w.h.p. at some  round $\tilde t=O\left(\log n\right)$
		the process reaches a   configuration $\genC^{(\tilde t)}$ such that:

	\begin{empheq}[left=\empheqlbrace]{align}
		&C_{\m}^{(\tilde t)}  \geq\left(\frac{1}{16}- {\frac \epsilon 8}\right)\frac{n}{R(\genC^{(\tilde t)}) }
			\label{hp:ub_phase2_clower}\\
		&R(\genC^{(\tilde t)})  \leq \cdist \cdot \left( 1+o\left( 1 \right) \right)
			\label{hp:ub_phase2_Rvalue}\\
		&C_{\m}^{(\tilde t)}  \geq  \left(1+\frac{\alpha}{2}\right)
			\cdot C_{i}^{(\tilde t)} \mbox{ for any color } i\neq\m
			\label{hp:ub_phase2_bias}\\
		&\frac{ C_{\m}^{(\tilde t)}+2Q^{(\tilde t)} }{n} > 1+\frac{\epsilon^2}{4}
			\label{hp:ub_phase2_growthfactor}
	\end{empheq}

	\end{lemma}

		\proof
			We prove one claim at a time. 

			\noindent
		\proofof{\eqref{hp:ub_phase2_clower}}
			Let $\tilde \epsilon$ be any positive constant in $ (\frac {\epsilon}2,\epsilon)$. 

			\noindent
			Two cases may arise. If $\bar c_{\m}>\left(\frac{1}{4}-\frac{\tilde \epsilon}{2}\right)\cdot n$,
			by applying the Chernoff bound \eqref{eq:CB_lower_multip} on the expected value of $C_{\m}^{(1)}$ 
			and using   \eqref{fa:ranger},	
			it is easy to see that w.h.p. 
			\[
				C_{\m}^{(1)} \geq \left(\frac{1}{16}-{\frac \epsilon 8}\right)n \geq
				\left(\frac{1}{16}- {\frac \epsilon 8}\right)\frac{n}{R (\genC^{(1)})}
			\]
			Assume now $\bar c_{\m} \leq\left(\frac{1}{4}-\frac{\tilde \epsilon}{2}\right)\cdot n$. 
			From Lemma \ref{lem:first_step_bounds} at round $t=1$
			we have w.h.p. 
			\[
				Q^{(1)} \geq {n}\left(1 - \frac{2} { \rratio(\bar \genc) } \right) 
				\geq {n}\left(1 - \frac{2  c_{\m}} { n } \right) \geq \frac{n}{2}+\tilde \epsilon \cdot n
			\]
			where we used that $\rratio(\bar \genc) \geq R(\bar \genc)=n/\bar c_{\m}$. 
			
			\noindent
			In the generic configuration $\genc$, as long as $q\geq \frac{n}{2}+\tilde \epsilon \cdot n$, from \eqref{eq:average_of_c} we have
			$$
				\medm
				\geq c_{\m} \cdot \left( \frac{1}{2}+\tilde \epsilon \right)
			$$
			thus, by applying  the Chernoff bound \eqref{eq:CB_lower_multip}, 
			we see that w.h.p. $C_{\m}$ grows exponentially fast. 

			\noindent
			It follows that 
			we can consider the first round	such that $\tilde t=O\left(\log n\right)$ and
			$Q^{(\tilde t)}<\frac{n}{2}+\tilde \epsilon\cdot n$. 
			This implies that 
			\[
				n-Q^{(\tilde t)}\geq\frac{n}{2}-\tilde \epsilon\cdot n
			\] 
			hence
			\[
			C_{\m}^{(\tilde t)}= \frac{n-Q^{(\tilde t)}}{R(\genC^{(\tilde t)})}  
					\geq  \frac{\frac{n}{2}-\tilde \epsilon\cdot n}{R(\genC^{(\tilde t)})}
			\]
			This proves  \eqref{hp:ub_phase2_clower}.

			\smallskip 
			\noindent
		\proofof{\eqref{hp:ub_phase2_Rvalue}}
			Observe that, since $\bar c_{\m} \geq \frac nk$, 
			then from \eqref{eq:average_of_c} and the Chernoff bound \eqref{eq:CB_lower_multip} 
			it holds w.h.p. that $C_{\m}^{(1)}=\omega(\log^{2}n)$.
			As we have already shown in the proof of Claim \eqref{hp:ub_phase2_clower}, 
			after the first round $C_{\m}$ grows exponentially until round $\tilde t$. 
			It follows that  we can repeatedly apply Lemma \ref{lem:R_any_step} and, 
			together with Lemma \ref{lem:R_first_step},
			we get w.h.p.  holds w.h.p. that 
			\begin{equation*}
				R ( \genC^{(\tilde t)} ) \leq \cdist\cdot\left(1+o\left(\frac{1}{\log n}\right)\right)^{\log n}
					   \leq \cdist\cdot\left(1+o\left(1\right)\right)
			\end{equation*}
			This proves   \eqref{hp:ub_phase2_Rvalue}.

			\smallskip
			\noindent
		\proofof{\eqref{hp:ub_phase2_bias}}
			Similarly to the previous Claim proof, the repeated application of Lemma \ref{lem:R_any_step} until round $\tilde t$ 
			and Lemma \ref{lem:inequality}  implies that w.h.p.
			\begin{align*}
				C_{\m}^{(\tilde t)} & \geq\left(1+\alpha\right)
					\cdot C_{i}^{(\tilde t)}\cdot\left(1-o\left(\frac{1}{\log n}\right)\right)^{\log n}\\
				 & =\left(1+\alpha\right)\cdot C_{i}^{(\tilde t)}\cdot\left(1-o\left(1\right)\right)
				 \geq\left(1+\frac{\alpha}{2}\right)\cdot C_{i}^{(\tilde t)}
			\end{align*}
			This proves   \eqref{hp:ub_phase2_bias}.

			\smallskip 
			\noindent
		\proofof{\eqref{hp:ub_phase2_growthfactor}}
			Since, by the definition of $\tilde t$, it holds $q^{(\tilde t -1)}\geq \frac n2 + \tilde \epsilon$, 
			then by Lemma \ref{lem:maj_drift} we get that  
			\begin{equation*}
				\expec{C_{\m}^{(\tilde t)}+2Q^{(\tilde t)}}{\genc^{(\tilde t -1)}}  \geq (1+\tilde\epsilon^2)\cdot n
			\end{equation*}
			Observe that $\expec{C_{\m}^{(\tilde t)}+2Q^{(\tilde t)}}{\genc^{(\tilde t -1)}}$ 
			can be written as the expected value of the sum of the following independent r.v.s:
			given a color configuration $\genc^{(\tilde t -1)}$, for each node $i$ 
			\begin{equation*}
				X_i = \left\{
					\begin{array}{ll}
						1 & \quad \text{if node $i$ is $\m$-colored at the next round},\\
						2 & \quad \text{if node $i$ is undecided at the next round}.
					\end{array} \right.
			\end{equation*} 
			Then   \eqref{hp:ub_phase2_growthfactor} is an easy application of the Chernoff bound \eqref{eq:CB_lower_multip}.
			\qed

\smallskip
From  the  state conditions 
achieved after the first round (see Lemma \ref{lem:first_step_bounds}), the next lemma shows that, 
within $O(\log n)$ rounds, 
the process w.h.p. reaches a configuration where 
$Q$ gets  very  close to $n/2$ and $C_{\m}$ is still relatively  small. In the next section, we will prove
(see Theorem \ref{thm:lowerbound})
that  this fact forces
the process to ``wait'' for  a  time period $\Omega\left(\mdist(\bar \genc)\right)$ before the plurality (re-)starts
to grow fastly. This   is the key ingredient of the  lower bound in Theorem \ref{thm:lowerbound}.

\begin{lemma}\label{lem:lb_phase2} 
 Let $k \leqslant \varepsilon \cdot (n/\log n)^{1/6}$ be the initial number of colors, where $\varepsilon > 0$ is a sufficiently small positive constant.
 Let   $\bar{\genc}$ be the initial color configuration and let $\genc^{(1)}$ be the  color configuration after the first round. If it holds that:
\begin{equation*}
	\frac{1}{2} \,  \frac{n}{ R(\bar{\genc})^2} \, \leq  c_{\m}^{(1)}  \leq \, 2 \,  \frac{n}{R(\bar{\genc})^2}
	\quad  \mbox{ and } \quad
	n \left(1 -  \frac{2   } { \rratio(\bar{\genc})}\right)
	\, \leq  q^{(1)}  \leq \, n \left(1 -  \frac{ 1 } { 2 \rratio(\bar{\genc})}\right)  
\end{equation*}
within the next $O(\log n)$ rounds there will be a round $\bar{t}$ such that  
$$
C^{(\bar{t})}_{\m} \leqslant \gamma \frac n{\mdist(\bar{c})} \ \mbox{and } \ \left| Q^{(\bar{t})} - \frac{n}{2} \right| \, \leq \, 2 \frac{\gamma^2}{\mdist(\bar{\genc})}
$$ 
  w.h.p., where $\gamma > 0$ is a sufficiently large constant.
\end{lemma}
\proof
First, we prove that if at an arbitrary round $t$ the number of undecided nodes is
$q = (1+\delta)(n/2)$ with $1/\mdist(\bar{\genc}) \leq \delta \leq 1 - (2\rratio(\bar{\genc}))^{-1}$,
then at the next round it holds that $Q' \leq \left(1 + \delta^2\right)(n/2)$  w.h.p.
Indeed, if  we replace  $q = (1+\delta)(n/2)$ in \eqref{eq:average_of_q},
we get that the expected value of $Q'$   at the next round is
\begin{eqnarray*}
	\medq
		& = & \frac{1}{n} \left( \left( (1+\delta)\frac{n}{2} \right)^2 + \left( (1+\delta)\frac{n}{2} \right)^2 
		- \sum_{j=1}^k \left(c_j\right)^2 \right) \\
	& = & \left(1 + \delta^2 \right) \frac{n}{2} - \frac{1}{n} \sum_{j=1}^k \left(c_j\right)^2
\end{eqnarray*}
Now observe that 
$$
\frac{1}{n} \sum_{j=1}^k \left(c_j\right)^2 
\geq \frac{1}{n} k \left( \frac{n-q}{k} \right)^2
= \frac{n }{4 k}(1-\delta)^2
\geq \frac{n}{4 k} \cdot \left( \frac{1}{2 \rratio(\bar{\genc})} \right)^2
\geq \frac{n}{16 k^3}
$$
where in the last inequality we used \eqref{le:ratio1}, that is $\rratio(\bar \genc) \leq {k}$.

\noindent
Therefore, since $Q'$ is a sum of independent Bernoulli r.v., 
from the Chernoff bound (Lemma~\ref{lemma:cbadditive} with $\lambda = 1/16k^3$) 
it follows that
\begin{eqnarray}\label{eq:superexp_decreasing_of_q}
	\probc{Q' \geq \left( 1 + \delta^2 \right) \frac{n}{2}}{\genc}
	& \leq & \exp\left( -\frac{n}{128 k^6} \right)\leq n^{-1/\left( 128 \varepsilon^6 \right)} 
\end{eqnarray}
where in the last inequality we used the hypothesis on $k$.

\smallskip
\noindent
Now we show that the number $Q$ of undecided nodes, while decreasing quickly, cannot jump over the whole interval
$$
\left[\frac{n}{2} - 2 \gamma^2 \frac{n}{\mdist(\bar{\genc})}, \; \frac{n}{2} + 2 \gamma^2 \frac{n}{\mdist(\bar{\genc})} \right]
$$ 
Observe that function $f(q) = q^2 + (n-q)^2$ has a minimum for $q = n/2$, so for any $q \geqslant n/2 + 2 \gamma^2 n /\mdist(\bar{\genc})$ it holds that $f(q) \geqslant f\left(n/2 + 2 \gamma^2 n /\mdist(\bar{\genc}) \right)$. Hence if at some round $t$ we have that $q \geqslant (n/2)\left(1 + 4 \gamma^2/\mdist(\bar{\genc})\right)$ and $c_1 \leqslant \gamma n /\mdist(\bar{\genc})$, in~(\ref{eq:average_of_q}) we get
\begin{eqnarray*}
\medq & \geqslant & \frac{1}{n} \left( \left( \frac{n}{2} + 2 \gamma^2 \frac{n}{\mdist(\bar{\genc})} \right)^2 + \left( \frac{n}{2} + 2 \gamma^2 \frac{n}{\mdist(\bar{\genc})} \right)^2 - \sum_{j=1}^k c_j^2 \right) \\
& = & \frac{n}{2} + 4 \gamma^4 \frac{n}{\mdist(\bar{\genc})^2} - \frac{1}{n} \sum_{j=1}^k \left(c_j\right)^2 \\
& \geqslant & \frac{n}{2} - \frac{1}{n} \sum_{j=1}^k \left(c_j\right)^2 = \frac{n}{2} - \frac{\left(c_1\right)^2 \mdist(\bar{\genc})}{n} \geqslant \frac{n}{2} - \gamma^2 \frac{n}{\mdist(\bar{\genc})}
\end{eqnarray*}
where in the last inequality we used that $c_1 \leqslant \gamma n /\mdist(\bar{\genc})$. Since $Q'$ is a sum of $n$ independent Bernoulli r.v., from Chernoff bound it follows that
\begin{eqnarray}\label{eq:q_not_too_below_nhalf}
	\probc{Q' \leqslant n/2 - 2 \gamma^2 n / \mdist(\bar{\genc}) }{ \genc}
& \leqslant & \exp \left( - 2 \gamma^2  \frac{n}{\mdist(\bar{\genc})^2} \right)
\leqslant \exp \left( - 2 \gamma^2  \frac{n}{k^2} \right) \nonumber\\
& \leqslant & \exp \left( - \Omega\left( n^{2/3} \right) \right)
\end{eqnarray}

\noindent
From \eqref{eq:superexp_decreasing_of_q}, we get that w.h.p. 
\begin{equation}\label{eq:undecsol}
Q^{(t)} \leq \left(1 + \delta^{2^t} \right) \frac{n}{2}
\end{equation}
\noindent
Hence, within 
\[
\log \left( \rratio(\bar{\genc}) \right)+ {O}(\log \log  \mdist(\bar{\genc})) 
\] 
rounds, the number $Q$ of undecided nodes will be below $(n/2) (1 + 4 \gamma^2 /\mdist(\bar{\genc}))$ w.h.p. Moreover, from (\ref{eq:q_not_too_below_nhalf}) it follows that in one of such  rounds we will have that
$$
\left| Q - \frac{n}{2} \right| \leq 2 \gamma^2/\mdist(\bar{\genc}) 
$$
w.h.p. It remains to show that, during this time, the plurality $C_{\m} $ does not increase from less $2n/R(\bar{\genc})^2$ to more than $\gamma n/\mdist(\bar{\genc})$.

\noindent
To simplify notation, let us define 
\begin{eqnarray*}
l & = & \log \left( \rratio(\bar{\genc}) \right) \\
L & = & \log \left( \rratio(\bar{\genc}) \right) + O(\log \log \mdist(\bar{c}))
\end{eqnarray*}

\noindent
From \eqref{eq:average_of_c} and~(\ref{eq:undecsol}) it follows that, 
as long as $c_{\m} \leqslant \gamma n / \mdist(\bar{\genc})$, 
the increasing rate of $C_{\m}$ at round $t$ is w.h.p. at most
\[
1 + \delta^{2^{t}} + \frac{\gamma}{\mdist(\bar{\genc})}
\]
For the first
$l$ rounds, we can bound the above increasing rate with $2$. 
Thus, after $l$ rounds we get that the plurality is $C_{\m} \leqslant 2 n/\mdist(\bar{\genc})$  w.h.p. As for the next $O\left(\log \log \mdist(\bar{\genc})\right)$
rounds, we have that the plurality is w.h.p. at most
\begin{eqnarray*}
2 \frac{n}{\mdist(\bar{\genc})} \cdot \prod_{t=l}^{L} \left( 1 + \delta^{2^t} + \frac{\gamma}{\mdist(\bar{\genc})} \right)
& \leq & 2 \frac{n}{\mdist(\bar{\genc})}
\cdot \exp\left(\sum_{t=l}^{L}
\left( \delta^{2^t} + \frac{\gamma}{\mdist(\bar{\genc})} \right) \right) \\
& \leq & 2 \frac{n}{\mdist(\bar{\genc})}
\cdot \exp\left( O(1) + \frac{\log\log \mdist(\bar{\genc})}{\mdist(\bar{\genc})} \right) \\
& \leqslant & \gamma \frac{n}{\mdist(\bar{\genc})} 
\end{eqnarray*}
where in the last inequality we need to choose $\gamma$ sufficiently large.
\qed

\medskip
\noindent
\textbf{Remark.} The two lemmas above refer to some rounds $\tilde{t},\bar{t} = O(\log n)$ in which the process lies in a state  satisfying  certain 
properties. We observe that our analysis does never  combine the two lemmas
and thus it does not require that $\tilde{t}=\bar{t}$, indeed the first lemma is used to get the upper bound while the second one to get the lower bound on the convergence time. However, it is possible to prove that there is in fact
a time interval (at the end of Phase 2)  where both claims of the lemmas hold w.h.p.

%% file: trunk/middleage.tex
\subsection{Second phase: \textit{Plateau} or \textit{Age of stability}}
This phase is characterized by a slow increase of $c_{\m}$, roughly
at a rate $1 + \Theta(1/\mdist(\bar \genc))$. This fact is formalized in the next lemma and it will be used
to derive the lower bound on the convergence time of the process in Theorem \ref{thm:lowerbound}.

\begin{lemma}\label{lem:lb_plateau}
Let $\bar{\genc}$ be the initial color configuration, let $k \leqslant \varepsilon \cdot (n/\log n)^{1/4}$ be the initial number of colors, where $\varepsilon > 0$ is a sufficiently small positive constant. If there is a round $\bar{t}$ such that
$$
\left| q^{(\bar{t})} - \frac{n}{2} \right| \, \leq \;
2 \gamma^2 \frac{n}{\mdist(\bar{\genc})}
\qquad \mbox{ and }\qquad
c_{\m}^{(\bar{t})} \leqslant \gamma (n/\mdist(\bar{\genc}))
$$
(where $\gamma$ is an arbitrary positive constant), 
then the plurality $C_{\m}$ remains smaller than $2 \gamma (n / \mdist(\bar{\genc}))$ 
for the next $\Omega(\mdist(\bar{\genc}))$ rounds w.h.p.
\end{lemma}
\proof
Let us define $\delta = q - n/2$ and let $\Delta'$ 
be the random variable $Q' - n/2$ in the next round. 
From (\ref{eq:average_of_c}) we get 
\begin{eqnarray}
\expec{\Delta'}{ \genc}
& = & \frac{1}{n} \left(2 \delta^2 - \sum_{j=1}^k \left(c_j\right)^2 \right) 
\label{eq:lb_av_delta}\\
\medi & = & 
\left( 1 + \frac{2 \delta + c_i}{n} \right) c_i \label{eq:lb_av_cm}
\end{eqnarray}
We now show that, if $\delta \in \left(- 2 \gamma^2 n/\mdist(\bar{\genc}), \, 2 \gamma^2 n/\mdist(\bar{\genc}) \right)$ 
and $c_{\m} \leq 2 \gamma \, n/\mdist(\bar{\genc})$, then the increasing rate of $C_{\m}$ is smaller than $(1 + \Theta(1/\mdist(\bar \genc)))$ w.h.p.
More precisely, we prove that
\[
\left\{
\begin{array}{rcl}
\left| \delta \right| & \leq & 2 \gamma^2 \frac{n}{\mdist(\bar{\genc})} \\[3mm]
c_{\m} & \leq & 2 \gamma \, \frac{n}{\mdist(\bar{\genc})}
\end{array}
\right.
\quad
\Longrightarrow
\quad
\left\{
\begin{array}{rcl}
\left| \Delta' \right| & \leq & 2 \gamma^2 \frac{n}{\mdist(\bar{\genc})} \\[3mm]
C_{\m}' & \leq & \left( 1 + \frac{2 \gamma(\gamma + 1) + 1}{\mdist(\bar{\genc})} \right) c_{\m}
\end{array}
\right.
\quad 
\mbox{ w.h.p. }
\]
As for the increasing rate of the plurality, from \eqref{eq:lb_av_cm} it follows that
\begin{eqnarray*}
\medm & = & \left( 1 + \frac{2 \delta + c_{\m}}{n} \right) c_{\m} \\
& \leq & \left( 1 + \frac{2 \gamma^2 n/\mdist(\bar{\genc}) + 2 \gamma n/\mdist(\bar{\genc})}{n} \right) c_{\m} 
= \left( 1 + \frac{2 \gamma(\gamma + 1)}{\mdist(\bar{\genc})} \right) c_{\m}
\end{eqnarray*}
Since $C_{\m}'$ can be written as a sum of 
$q + c_{\m} \leq n$ independent Bernoulli random variables, 
from Chernoff bound (Lemma \ref{lemma:cbadditive} with $\lambda = c_1 / (n \mdist(\bar{\genc}))$ it follows that
\begin{eqnarray}\label{eq:cmconc}
\probc{C_{\m} \geq \left(1 +\frac{2\gamma(1+\gamma)+1}{\mdist(\bar{\genc})} \right)c_{\m}}{ \genc}
& \leq & \exp\left(- \frac{2 \left(c_{\m}/\mdist(\bar{\genc})\right)^2}{n}\right) \nonumber \\
& \leq & \exp\left(- \frac{2 n}{9 k^4}\right)
\leq n^{-2 / (9 \varepsilon^4)}
\end{eqnarray}
where in the second inequality we used the fact that $c_{\m} \geq n -q/k \geq n/(3k)$ and $\mdist(\bar{\genc}) \leqslant k$, and in the last inequality we used the hypothesis $k \leq \varepsilon \cdot (n/ \log n)^{1/4}$.

\smallskip\noindent
As for $\expec{\Delta'}{ \genc}$, 
according to \eqref{eq:lb_av_delta}, we have the upper bound
\begin{equation}\label{eq:ubexpecdelta}
\expec{\Delta'}{ \genc} \leq 2 \frac{\delta^2}{n} 
\leq 8 \gamma^4 \frac{n}{(\mdist(\bar{\genc}))^2}
\leq \gamma^{2} \frac{n}{\mdist(\bar{\genc})^2}
\end{equation}
where in the first inequality we discarded the non-negative term $\sum_{j=1}^k \left(c_j\right)^2$, in the second inequality we have used  $| \delta | \leq 2 \gamma^2 n/\mdist(\bar{\genc})$, and in the third one we simply assumed that $\mdist(\bar{\genc})$ is a sufficiently large constant, namely $\mdist(\bar{\genc}) \geqslant 8 \gamma^2$.

\noindent
On the other hand, we have the lower bound
\begin{eqnarray}\label{eq:lbexpecdelta}
\expec{\Delta'}{\genc} 
& = & \frac{1}{n} \left( 2 \delta^2 - \sum_{j=1}^k \left(c_j\right)^2 \right) 
\geq  - \frac{1}{n} \sum_{j=1}^k \left(c_j\right)^2 \nonumber \\
& \geq & - \frac{k}{n} \left( \frac{n - q}{k} \right)^2
\geq - \frac{4}{9} \cdot \frac{n}{k} \geq - \frac{4}{9} \cdot \frac{n}{\mdist(\bar{\genc})}
\end{eqnarray}
From the first to the second line we used the fact that 
all $c_j$'s are smaller than $n - q$. Then 
we used the fact that $q$ is close to $n/2$, so $n - q$ is smaller than, say, $(2/3) n$. Finally we used the fact that $k \geq \mdist(\bar{\genc})$. 

\noindent
Hence, from \eqref{eq:ubexpecdelta} and \eqref{eq:lbexpecdelta} we get
$$
- \frac{4}{9} \frac{n}{\mdist(\bar{\genc})}
\; \leqslant \, \expec{\Delta'}{ \genc} \, \leqslant \; 
\gamma^2 \frac{n}{\mdist(\bar{\genc})}
$$

\noindent
Since $\Delta' = Q' - n/2$ can be written as a sum of $n$ independent random variables taking values $\pm 1/2$, from the appropriate version of Chernoff bound it thus follows that
\begin{equation}\label{eq:deltaconc}
\probc{\Delta' \notin \left(-2 \gamma^2 \frac{n}{\mdist(\bar{\genc})},\, 2 \gamma^2 \frac{n}{\mdist(\bar{\genc})} \right)}{\genc }
\leq \exp\left(-\Omega\left(\frac{n}{\mdist(\bar{\genc})^2} \right) \right)
\leq \exp \left( -\Omega\left( n^{1/2} \right) \right)
\end{equation}
where in the last inequality we used again the fact that $\mdist(\bar{\genc}) \leqslant k \leqslant \varepsilon \left(n/\log n\right)^{1/4}$.

\smallskip\noindent
In order to formally complete the proof, let us now define event $\mathcal{E}_t$ as follows
$$
\mathcal{E}_t = \mbox{``} |\Delta^{(t)}| \leq 2 \gamma^2 \frac{n}{\mdist(\bar{\genc})} 
\, \mbox{ and } \, 
C_{\m}^{(t)} \leq \left(1 + \frac{2\gamma(1+\gamma) + 1 }{\mdist(\bar{\genc})} \right)^t 
\cdot \gamma \frac{n}{\mdist(\bar{\genc})} \mbox{''} 
$$
Observe that 
$$
\left( 1 + \frac{2\gamma(1+\gamma)+1}{\mdist(\bar{\genc})} \right)^t \leq 2 \ \mbox{ for } \ t \leq \frac{1}{4\gamma(1+\gamma)} \cdot \mdist(\bar{\genc})
$$
Hence, if we set $T = \left\lfloor \frac{1}{4\gamma(1+\gamma)}\mdist(\bar{\genc}) \right\rfloor$, from~\eqref{eq:cmconc} and~\eqref{eq:deltaconc} it follows that, for every $j \in [\bar{t}, \bar{t} + T]$, we get
$$
\probc{\mathcal{E}_j}{ \cap_{i=1}^{j-1}\mathcal{E}_{i}} \geq (1 - n^{-c}) 
$$
for a positive constant $c$ that we can make arbitrarily large. 
Thus, starting from the given color configuration $\genc^{(\bar{t})}$, 
the probability that after $T$ rounds
the plurality $C_{\m}^{(\bar{t}+T)}$ is at most $2 \gamma n/\mdist(\bar{\genc})$ is
\begin{eqnarray*}
\probc{C_{\m}^{(\bar{t}+T)} \leq 2 \gamma \frac{n}{\mdist(\bar{\genc})}}{\genc^{(\bar{t})}} 
& \geq & \probb{\bigcap_{j = \bar{t}}^{\bar{t}+T} \mathcal{E}_j }= \prod_{j=\bar{t}}^{\bar{t}+T} 
\probc{\mathcal{E}_j }{ \bigcap_{i=\bar{t}}^{j-1} \mathcal{E}_i } \geq \\
& \geq & \left(1 - n^{-c} \right)^{T} \geq 1 - T n^{- c} \geq 1 - n^{-\Omega(1)}
\end{eqnarray*}
\qed

\begin{theorem} \label{thm:lowerbound}
Let $\bar{\genc}$ be the initial color configuration. If the initial number of colors is $k \leqslant \varepsilon \cdot (n/\log n)^{1/6}$, where $\varepsilon > 0$ is a sufficiently small positive constant, then the convergence time of the \threestate\ is $\Omega(\mdist(\bar{\genc}))$ w.h.p.
\end{theorem}
\proof
From Lemma~\ref{lem:first_step_bounds} and Lemma~\ref{lem:lb_phase2} it follows that there is a round $\bar{t}$, within the first $O(\log n)$ rounds, such that the process lies in a color configuration $\genc^{(\bar{t})}$ where the number of undecided nodes is $\left| Q^{(\bar{t})} - n/2 \right| \leq 2 \gamma^2/\mdist(\bar{\genc})$ and the plurality is $C_{\m}^{(\bar{t})} \leq \gamma (n/\mdist(\bar{\genc}))$ w.h.p., where $\gamma$ is a sufficiently large constant. 
From Lemma~\ref{lem:lb_plateau}, it then follows that the plurality $C_{\m}$ remains smaller than $2 \gamma (n / \mdist(\bar{\genc}))$ for the next $\Omega(\mdist(\bar{\genc}))$ rounds.\qed

\smallskip\noindent
There is, however, a ``positive'' drift  for the plurality working in this ``long'' phase as well: this minimal drift 
(see the next lemma) allows the process to reach a state (representing the end of this phase)
by which the plurality can re-start to grow fast (this phase-completion state
is formalized in Lemma \ref{lem:ub_plateau}).
 
	\begin{lemma}[Minimal Drift]
		\label{lem:phases_drift}
		Let $k=o\left(\sqrt{\frac{n}{\log n}}\right)$ and let $\epsilon \in (0,\frac{1}{2})$ be an arbitrarily small positive constant. 
		Given a color configuration $\genc$ such that
		\begin{equation*}
			\left\{
			\begin{array}{l}
				c_{\m} \geq \beta \cdot \frac{n}{R (\genc)} \mbox{ for some constant } \beta >0\\
				c_{\m}\geq \left( 1+\alpha \right)c_i \mbox{ for some constant } \alpha>0 \mbox{ and any } i\neq \m
			\end{array}
			\right.
		\end{equation*}
		w.h.p. it holds either 
		\begin{equation*}
			R ( \genC' )  \leq 1+\frac{\epsilon}{3} \text{ and } Q' \leq \epsilon n
		\end{equation*}
		or
		\[
			\frac{ C_{\m}'+2Q'}{n}
			\geq  1+\Omega\left( \frac{1}{R \left( \genc \right)} \right)
		\] 
	\end{lemma}
		\proof
			First, let us derive a lower bound on $C_{\m}'+2Q'$ that holds w.h.p. 

			\noindent
			By Lemma \ref{lem:maj_drift}
			\[
				\expec{{ C_{\m}'+2Q'}}{\genc} =n\cdot \left( 1+\Gamma(\genc) \right)
			\] 
			where 
			\[
				\Gamma(\genc)=\left( 1-\frac{ c_{\m}+2q}{n} \right)^{2} + 2\left( 1-\gamma \right)\left( R (\genc)-1 
				\right)\left(\frac{ c_{\m} }{n}\right)^{2}
			\] 
			with $\gamma=\left( 1+\alpha \right)^{-1}$.
			As in the proof of Lemma \ref{lem:ub_phase2}, observe that $\expec{C_{\m}'+2Q'}{\genc}$ 
			can be written as the expected value of the sum of the following independent r.v.s:
			given $\bar \genc$, for each node $i$ 
			\begin{equation*}
				X_i = \left\{
					\begin{array}{ll}
						1 & \quad \text{if node $i$ is $\m$-colored at round $t+1$},\\
						2 & \quad \text{if node $i$ is undecided at round $t+1$}.
					\end{array} \right.
			\end{equation*} 
			Thus, we can apply the Chernoff bound \eqref{eq:CB_lower_multip} 
			to them and get that w.h.p.  
			\begin{equation}
				C_{\m}'+2Q'\geq n\cdot \left( 1+\Gamma(\genc) \right)
					\left( 1-O\left( \sqrt{\frac{\log n}{n}} \right) \right)
				\label{eq:maj_drift_proof}
			\end{equation}
			
			\noindent
			Let us analyze \eqref{eq:maj_drift_proof} when $R (\genc) > 1 + \frac{\epsilon}{4}$ or $Q' > \frac 34 \epsilon n$.

			\noindent
			If $R (\genc) > 1 + \frac{\epsilon}{4}$ we have that 
			\begin{align}
				&\Gamma(\genc)\geq 2\left( 1-\gamma \right)\left( R (\genc)-1 \right)
					\left(\frac{ c_{\m} }{n}\right)^{2} \geq \nonumber\\
				&\geq 2\left( 1-\gamma \right)\left( 1-\frac{1}{R (\genc)} \right)
				R (\genc)\cdot  \left(\frac{\beta}{R (\genc)}\right)^{2} >
				\frac { \alpha \epsilon \beta^{2}} { 2 (1+\alpha)(1+\epsilon/4)}\frac1{R (\genc)}
				\label{eq:phases_drift_R}
			\end{align}
			On the other hand, if $R (\genc) \leq 1 + \frac{\epsilon}{4}$ then
			\begin{align*}
				c_{\m} &= \frac{n-q}{R (\genc)} \geq  \frac{n-q}{ 1 + \epsilon/4} 
					\geq  (n-q)( 1 - \epsilon/4) \geq n-q-\frac{\epsilon}4 n
			\end{align*}
			hence, if it also holds that $q > \frac{3}{4}\epsilon n$, the latter inequality implies that
			\begin{equation*}
				1-\frac{ c_{\m}+2q}{n} \leq \frac{\epsilon}4-\frac{q}n \leq -\frac{\epsilon}{2}
			\end{equation*}
			that is
			\begin{equation}
				\label{eq:phases_drift_q}
				\Gamma(\genc)\geq \left( 1-\frac{ c_{\m}+2q}{n} \right)^{2}  \geq \frac{\epsilon^{2}}{4}
			\end{equation}

			\noindent
			Therefore, if $R (\genc) > 1 + \frac{\epsilon}{4}$ or $q > \frac{3}{4}\epsilon n$, 
			then using \eqref{eq:phases_drift_R}, \eqref{eq:phases_drift_q} 
			and the given upper bound on the value of $R (\genc)$, 
			from \eqref{eq:maj_drift_proof} we get
			\begin{align*}
				\frac{ C_{\m}'+2Q'}n &\geq \left( 1+\Gamma(\genc) \right)
					\left( 1-O\left( \sqrt{\frac{\log n}{n}} \right) \right) \geq \\
				&\geq \left( 1+\frac{\sigma}{R (\genc)} \right) \left( 1-O\left( \sqrt{\frac{\log n}{n}} \right) \right)
					\geq \left( 1+\frac{\sigma}{2R (\genc)} \right) 
			\end{align*}
			where
			\[
				\sigma=\min\left\{  \frac{\epsilon^{2}}{4}R (\genc), \frac{ \alpha \epsilon \beta^{2}}{2 (1+\alpha)(1+\epsilon/4)} \right\}
			\]
			
			\noindent
			It remains to show that if $R (\genc) \leq 1 + \frac{\epsilon}{4}$ and $q \leq \frac{3}{4}\epsilon n$
			then w.h.p. $R (\genC') \leq 1 + \frac{\epsilon}{3}$ and $Q' \leq \epsilon n$.

			\noindent
			In order to do so, observe that
			\begin{equation*}
				\sum_{i\neq \m} c_i = {( R (\genc)-1 )}{c_{\m}} \leq  \frac{\epsilon}{4}n
			\end{equation*}
			It follows that
			\begin{align*}
				\medq &= \frac{q^2+\sum_{i\neq j} c_ic_j}{n} 
					\leq \frac{q^2+2c_1\sum_{j\neq \m} c_j 
					+ \sum_{i\neq \m} c_i\sum_{j\neq \m}c_j}{n} \leq \\
					&\leq \left(\frac{3}{4}\epsilon\right)^{2} n + \frac{\epsilon}2 c_{\m}+\frac{\epsilon^2}{16}n
			\end{align*}
			Thanks to the Chernoff bound \eqref{eq:CB_upper_multip} and since $\epsilon < \frac{1}{2}$, 
			the previous inequality implies that w.h.p. $Q'\leq \epsilon n$.
			As for $R (\genC')$, by applying Lemma \ref{lem:R_any_step} and using the Chernoff bound \eqref{eq:CB_upper_multip}, 
			we get that w.h.p. $R(\genC')\leq 1+\frac{\epsilon}3 $, concluding the proof.
		\qed

	\begin{lemma}\label{lem:ub_plateau}
		Let $k  = O(  (n/ \log n)^{1/4})$ and let $\epsilon>0$ be an arbitrarily small constant. 
		If the process is in a color configuration $\genc^{(\tilde t)}$ 
		that satisfies the following conditions:
		\begin{empheq}[left=\empheqlbrace]{align}
			&\frac{ c_{\m}^{(\tilde t)}+2q^{(\tilde t)} }{n} =  1+\Omega\left( \frac{1}{R (\genc^{(\tilde t)})} \right)
				\label{hp:ub_plateau_growthfactor} \\
			&c_{\m}^{(\tilde t)}  \geq  \frac{1}{17} \frac{n}{R (\genc^{(\tilde t)})}
				\label{hp:ub_plateau_clower} \\
				&R (\genc^{(\tilde t)}) = O(\mdist(\bar \genc))
				\label{hp:ub_plateau_Rvalue}\\
			&c_{\m}^{(\tilde t)}  \geq  \left(1+{\alpha}\right) 
				\cdot c_{i}^{(\tilde t)} \mbox{ for some constant } \alpha>0 \mbox{ and for any color } i\neq\m
				\label{hp:ub_plateau_bias}
		\end{empheq}
		then, after $T=O\left(\mdist(\bar \genc)\cdot\log n\right)$ rounds, the process is w.h.p. in a color configuration 
		$\genC^{(\tilde t + T)}$ such that 
		\begin{empheq}[left=\empheqlbrace]{align*}
			&C_{\m}^{(\tilde t + T)}  \geq  \frac{1}{17} \frac{n}{R(\genC^{(\tilde t+T)})}\\
			&R(\genC^{(\tilde t+T)}) \leq 1+\frac {\epsilon}3\\
			&Q^{( t +T)} \leq \epsilon n\\
			&C_{\m}^{(\tilde t + T)}  \geq  \left(1+{\alpha}\right) 
				\cdot C_{i}^{(\tilde t + T)}(1-o(1)) \mbox{ for any color } i\neq\m
		\end{empheq}
	\end{lemma}
		\proof
			First, we show that, if we start in a color configuration $\genc$ satisfying properties 
			\eqref{hp:ub_plateau_growthfactor}, \eqref{hp:ub_plateau_clower}, 
			\eqref{hp:ub_plateau_Rvalue} and \eqref{hp:ub_plateau_bias}, 
			then w.h.p.  $\genC'$ still satisfies the conditions
			\eqref{hp:ub_plateau_clower}, \eqref{hp:ub_plateau_Rvalue} and \eqref{hp:ub_plateau_bias}.

			\noindent
			Using the Chernoff bound \eqref{eq:CB_lower_multip} and conditions \eqref{hp:ub_plateau_clower} 
			and \eqref{hp:ub_plateau_growthfactor}, we get that w.h.p. 
			\[
				C_{\m}' \geq  \frac{ c_{\m}^{(\tilde t)}+2q^{(\tilde t)} }{n} c_{\m}
					\left( 1- O\left( \sqrt{\frac{\log n}{\medm}} \right)\right)
					= \left(1+\Omega\left( \frac{1}{R (\genc)} \right) \right)c_{\m}
					\geq  \frac{1}{17} \frac{n}{R{(\genc)}}
			\]
			In the first equality, we used that \eqref{hp:ub_plateau_growthfactor} and \eqref{hp:ub_plateau_clower}
			together imply that $\medm \geq c_{\m}\geq  \frac{1}{17} \frac{n}{R (\genc)} \gg \frac 1{R (\genc)}$ w.h.p., 
			thus proving that  $\genC'$ also satisfies Condition
			\eqref{hp:ub_plateau_clower} w.h.p.
			Moreover, Condition \eqref{hp:ub_plateau_clower} allows us to apply Lemma \ref{lem:R_any_step} to get that w.h.p. 
			\begin{gather*}
				C_{\m}'  \geq  \left(1+{\alpha}\right) \cdot C_{i}'
				\cdot \left(1-O\left(\left({\log n}/{\medm}\right)^{1/2}\right)\right)
				\quad
				\text{ and }
				\quad
				R{(\genC')}<R{(\genc)} 
					\cdot \left(1+O\left( \left( {\log n}/{\medm} \right)^{1/2} \right)\right)
			\end{gather*} 
			proving that w.h.p. $\genC'$ satisfies the hypotheses 
			\eqref{hp:ub_plateau_Rvalue} and \eqref{hp:ub_plateau_bias}.

			\noindent
			Now, by  Lemma \ref{lem:phases_drift} and \eqref{hp:ub_plateau_Rvalue}, it follows that w.h.p. either 
			$R(\genC') \leq 1+ \frac {\epsilon}3$ and $Q'\leq \epsilon n$ (in which case, we have done), 
			or it holds w.h.p. that
			\begin{equation*}
				\frac{ C_{\m}'+2Q' }{n} =  1+\Omega\left( \frac{1}{R (\genc)} \right) = 1+\Omega\left( \frac{1}{\mdist (\bar \genc)} \right)
			\end{equation*}
			In the latter case, $\genC'$ satisfies also Condition  
			\eqref{hp:ub_plateau_growthfactor} and the above argument can be iterated again.
			In particular, \eqref{hp:ub_plateau_growthfactor} implies that after
			$T=\Omega(\mdist(\bar \genc)\log n)$ further rounds w.h.p. we have
			\[ 
				C_{\m}^{(\tilde t + T)}
					= \left( 1+\Omega\left( \frac{1}{\mdist (\bar \genc)} \right)\right) 
					c_{\m}^{(\tilde t +T-1)}  
					= \dots
					= \left( 1+\Omega\left( \frac{1}{\mdist (\bar \genc)} \right)\right)^{T} 
					c_{\m}^{(\tilde t)} = n-o(n)
			\]
			and thus
			\begin{equation*}
				R (\genC^{( \tilde t+T )})-1 = \frac{\sum_{i\neq \m}C_{i}^{(\tilde t +T)}}{C_{\m}^{(\tilde t+T)}}
				\leq \frac {\epsilon}3 \text{ and } Q^{(\tilde t +T)}\leq \epsilon n
			\end{equation*}

		\qed

%% file: trunk/supremacy.tex
\subsection{Third phase: \textit{From plurality to totality}}
The next theorem  connects the results achieved in the 
previous sections into a consistent picture, establishing an upper bound on  
the  overall  convergence time of the process. Its proof also highlights 
the main features of the final phase, during which plurality turns 
into totality of the agents at an exponential rate. 
	
\begin{theorem}
\label{thm:upperB} 
Let $k=O\left( (n/\log n)^{1/3} \right)$ 
and let $\bar c$ be any initial configuration such that for any $i\neq \m$  
$c_{\m}\geq\left(1+\alpha\right)\cdot c_{i}$ holds, where
$\alpha$ is an arbitrarily small positive constant. 
Then, w.h.p. after at most $T=O\left(\cdist\cdot\log n\right)$ time steps
all agents support the initial plurality color.
\end{theorem}
\proof
Let $\epsilon>0$ be an arbitrarily small positive constant.
Thanks to  Lemma \ref{lem:ub_phase2}, 
we can assume that at some time $\tilde t=O(\log n)$ the process w.h.p. 
reaches a configuration $\genC ^{(\tilde t)}$ where

\begin{empheq}[left=\empheqlbrace]{align*}
 	&\frac{ C_{\m}^{(\tilde t)}+2Q^{(\tilde t)} }{n} =  1+\Omega\left( \frac{1}{R (\genc^{(\tilde t)})} \right)\\ 
	&C_{\m}^{(\tilde t)}  \geq  \frac{1}{17} \frac{n}{R (\genc^{(\tilde t)})}\\
 	&R (\genc^{(\tilde t)}) = O(\cdist)\\
 	&C_{\m}^{(\tilde t)}  \geq  \left(1+{\alpha}\right) 
		\cdot c_{i}^{(\tilde t)}(1-o(1)) \mbox{ for any color } i\neq\m
\end{empheq}

Assuming $\genc ^{(\tilde t)}$, Lemma \ref{lem:ub_plateau} determines the kick-off 
condition for a  new phase in which both the undecided and the 
non-plurality  color communities decrease exponentially fast. In particular, it implies that w.h.p.,
within $O(\cdist \log n)$ further rounds, the process reaches a configuration $\genC ^{(t_{\End})}$ 
such that the following properties hold:

\begin{empheq}[left=\empheqlbrace]{align}
	&C_{\m}^{(t_{\End})}  \geq  \frac{1}{17} \frac{n}{R (\genc^{(t_{\End})})} \label{item:upperB_C}\\
	&C_{\m}^{(t_{\End})}  \geq  \left(1+{\alpha}\right)
	       \cdot C_{i}^{(t_{\End})}(1-o(1)) \mbox{ for any color } i\neq\m
	\label{item:upperB_bias}\\
	&R (\genc^{(t_{\End})}) \leq 1+\frac{\epsilon}3\label{item:upperB_R}\\
	&Q^{t_{\End}} \leq {\epsilon}n\label{item:upperB_q}
\end{empheq}

\noindent
Now, we show that starting from any configuration satisfying the conditions
above, any community (including the undecided) other than the plurality 
decreases exponentially fast until disappearance.
To this aim, let $\psi=\sum_{i\neq \m} c_i+q$ and, as usual, let $\Psi'$ be the r.v. 
associated to the value of $\psi$ at the next time step.
We prove that the following holds in any round following $t_{\End}$:
i) w.h.p., both $Q$ and $\sum_{i\neq \m} C_i$ are bounded by quantities
that decrease by a constant factor, so that at any time following $t_{\End}$, 
$\Psi$ is (upper) bounded by a quantity that decreases exponentially fast,
thus $C_{\m}=n-\Psi$ is (lower) bounded  by an increasing quantity; ii) properties \eqref{item:upperB_bias}, 
still holds. In the rest of this 
proof we assume $\epsilon < 1/3$, which is consistent with the assumptions
of Lemma \ref{lem:ub_plateau}.

\noindent
To begin with, 
note that Property \eqref{item:upperB_R} implies $\sum_{i\neq \m} c_i \leq \frac{\epsilon}{3}n$,
so that 
\[
	\sum_{i\neq j}c_i\cdot c_j\leq
		2c_{\m}\sum_{j\neq 1}c_i
		+\sum_{i\neq \m}c_i \sum_{j\neq \m} c_j
		\leq  \left(\frac{2}{3}\epsilon + \frac{\epsilon^2}{9}\right)n^2
\]
Therefore, properties \eqref{item:upperB_R} and \eqref{item:upperB_q} together imply
\begin{align}
	&\medq
		= \frac{\left( q \right)^2
		+\sum_{i\neq j}c_i\cdot c_j}{n} 
		\leq \left( \epsilon^2 + \frac{2}{3}\epsilon + \frac{\epsilon^2}{9} \right)n
		< \frac 34 \epsilon n 
		\label{eq:upperB_nonplur_decrease}\\
	&\expec{\sum_{i \neq \m}C_{i}'}{\genc }
		= \sum_{i\neq \m} \left( c_i\frac{c_{i}+2q}{n}\right) 
		\leq  \frac 13\left( \frac 13 + 2 \right)\epsilon^2 n
		= \frac 79 \epsilon^2 n < \frac{7}{27}\epsilon
	\label{eq:upperB_q}
\end{align}
where we use the assumption that $\epsilon < 1/3$.
At this point, we can use the Chernoff bound \eqref{eq:CB_upper_multip} to show that 
\eqref{eq:upperB_nonplur_decrease} and \eqref{eq:upperB_q} hold w.h.p. 
(up to a multiplicative factor $1+o(1)$). This proves that w.h.p., both
$Q$ and $\sum_{i \neq \m}C_{i}$ (and hence $\Psi$) decrease by a constant
factor in a round\footnote{In fact, a more careful analysis,
unnecessary to prove our result, could use \eqref{eq:upperB_q} to show that
$\sum_{i \neq \m}C_{i}$ decreases superexponentially fast.}.
It remains to observe that, when $ q$ and/or $\sum_{i\neq \m}c_{i}$ become $O(\log n)$, 
an application of the Chernoff bound \eqref{eq:CB_absolute} shows that
w.h.p., they remain below this value in the subsequent rounds.
This completes the proof of i).
Moreover, since $C_{\m}'=n-\Psi'$, i) implies that $C_{\m}'$ is lower bounded by an increasing quantity w.h.p.
Additionally, property \eqref{item:upperB_C} and i) just proved, together
with property \eqref{item:upperB_bias}, imply the assumptions of  
Lemma \ref{lem:R_any_step}, allowing us to show that w.h.p. property
\eqref{item:upperB_bias} still holds at the end of next round as well.

\noindent
As a consequence, we have that in at most $\tau=O(\log n)$ rounds w.h.p. we reach a color configuration $\bar C^{(t_{\End}+\tau)}$ such that
$Q^{(t_{\End}+\tau)} + \sum_{i\neq \m}C_{i}^{(t_{\End}+\tau)} =O(\log n)$. 

\noindent
Finally, we can apply Markov's inequality on the value of $\sum_{i\neq \m}C_{i}^{(t_{\End}+\tau)} $
to show that at the next round w.h.p. all color communities except for the plurality one disappear.
\qed

%% file: trunk/expanders.tex
\section{The \threestate\ on expander graphs} \label{secmain:expanders}

The \threestate\  can be adapted to   compute
 plurality consensus  on the class of $d$-regular  expander  graphs \cite{HLW06} (where $d$ is   the degree of
 the  nodes)
 by paying  only a polylogarithmic extra-cost in terms of local memory and time.  
 
 The simple idea is to simulate the
 (uniform) random sampling of  neighbor  colors by the use of $n$ agent's tokens, each of them
  running a (short) random-walk over the graph.
  
 It is well known~\cite{lpwAMS08} that in every  $d$-regular expander   $G(V,E)$   
   a lazy random walk   has a uniform stationary distribution. Moreover,  
    it is \emph{rapidly mixing}, i.e., its    mixing time is $\bar t = O(\log(1/\epsilon)\log n)$ where $\epsilon$ 
 is the desired  bound on 
 the    total variation distance.
Formally,
let $\tm{G}{\epsilon}$ be the first round 
such that the total variation distance between the lazy simple random walk starting at an arbitrary node and the uniform distribution is smaller than $\epsilon$, i.e., 
$$
\tm{G}{\epsilon} = \inf \{ t \in \mathbb{N} \,:\, \| P^t(u, \cdot) - \pi\| \leqslant \epsilon \ \mbox{ for all } u \in V\}
$$
Notice that for any $\epsilon > 0$ it holds that (see e.g. (4.36) in~\cite{lpwAMS08})
\begin{equation}\label{eq:tvfall}
	\tm{G}{\epsilon} \leqslant \log (1/\epsilon) \tm{G}{1/(2e)}
\end{equation}

\smallskip
\noindent
\textbf{The modified \threestate . }
The modified dynamics works   in synchronous \emph{phases}, each 
of them consisting of exactly $2 \tau$ rounds (the suitable 
value for $\tau$ will be defined later). During the first $\tau$ 
rounds a {\em forward process} takes place: Every node sends a token performing a random walk of at least $\bar{t}$-\emph
{hops} and thus sampling a random color. In 
the next $\tau$ rounds we have a {\em backward process}: Every 
token is sent back to its source by ``reversing'' the path 
followed in the forward process.

If we were in the \local\ model~\cite{P00}, where each agent can 
communicate with all its neighbors in one round, each phase of 
the above protocol would last exactly $2 \bar{t}$ rounds. 
In the \gossip\ model~\cite{CHKM12},  each agent can instead 
activate only one (bidirectional) link per round. 
Moreover, since we want \emph{messages of limited size}, 
we assume that through each direction of an active link
only one token can be transmitted. 

We further assume that nodes enqueue 
tokens with a \emph{FIFO} 
policy, breaking ties arbitrarily. 
The random walk performed by
a token will thus likely require more than $\bar{t}$ rounds to perform (at least)
$\bar{t}$ hops of the random walk,  depending on the \textit{congestion}, i.e. 
the maximum number of tokens enqueued in a  node during  a round. 
We thus need to bound the maximal congestion and use this bound 
(together with $\bar{t}$) to suitably set  the right value for  
$\tau$ (valid for all tokens), so that every  token (i.e., the corresponding 
random walk) is w.h.p.  ``mixed'' enough. 
Finally, 
at time $2\tau$ each agent contains exactly its own token, and updates
its color according to the \threestate. 
After that, a new phase starts, and the process iterates.
Further important details and remarks about this modified dynamics:

	\begin{itemize}
	
	\item  During  the forward process, 
	every token records  the link labels of its random-walk   and 
	each node  records, for any   round,  the (local) 
	link label it has used (if any) to send  a token at that   round. Thanks to  this   information, 
	every node can easily perform the backward process  of the phase: 
	At every round of this process, 
	each node knows (if any) the neighbor it must contact to receive the right token back\footnote{Recall that in  the 
	\gossip\ model~\cite{CHKM12}, agents can indeed contact one \emph{arbitrary} neighbor per round.}. 
	Notice that, since the backward process is perfectly specular to the forward one, 
	the congestion is the same in both phases.
	Hence,  both node memory and token message require 
	$\Theta(\tau \log d)$ bits to  perform the phase.   
	 
	\item
	 By setting a suitable value for $\tau$, every  token  will w.h.p. perform at least  $\bar{t}$ 
	hops (some tokens may perform more hops than others). Thanks to the rapidly-mixing property, 
	the color reported to the sender is chosen nearly uniformly at random, i.e.,
	each agent has probability $1/n \pm \epsilon$ to be sampled (our analysis works setting $\epsilon  = O(1/n^2)$).
	
	\end{itemize}

In the next paragraph, we give our analysis of the node congestion. This analysis results into a concentration upper bound on the maximal node congestion during a phase of the  protocol.
As described above, this bound is crucial to set the value of 
  $\tau$ valid for  all random walks in every phase.

\medskip
\noindent
\textbf{Node congestion analysis.} 
The parallel random walks yield variable token queues in the nodes.
For each node $u \in [n]$, and for every round $t\in [2\tau]$ of the phase, we consider the
r.v. $\Q_{t,u}$ defined as the number of tokens in $u$ at round $t$ of any phase of the modified dynamics.
In the next lemma we prove a useful bound on 
the maximal congestion in a phase of length $2\tau$.

\begin{lemma}\label{lemma:congestion}
	Consider a phase of length $2 \tau \geqslant 1$ of the
	 above protocol on a $d$-regular graph $G = (V,E)$. Let $u \in V$ be any node and let $t$ be any round of the phase. Then, for any constant $c>0$, it holds that 
	$$
	\Prob{}{\max_{1 \leqslant t \leqslant 2\tau} \Q_{t,u} \leqslant \max \left\{ \sqrt{ 2 c \tau \log n}, \, 3 c \log n \right\}} \geqslant 1 - \frac{(2\tau)^2}{n^{c/3}}
	$$
\end{lemma}
\proof 
Consider the number
$Y_t$ of tokens received by a fixed node $u$ at round $t$ (for brevity's sake, we will omit index $u$
in any r.v.). 
Then we can write 

\[ Y_t = \sum_{i\in [d]} X_{i,t} \] where 
$X_{i,t}= 1$ if the $i$-th neighbor of $u$ sends a token to $u$ and $0$ o.w..
Observe (again) that the r.v.s $X_{i,t}$
are not mutually independent. However, the crucial fact is that, for any $t$ and any $i$, it holds 
$\Prob{}{X_{i,t} = 1} \leqslant 1/d$, \emph{ regardless the state of the system (in particular,
independently of the value of the other r.v.s)}. 

\noindent
So, if we consider a family $\{\hat X_{i,t} \,:\, i \in [d] \; t \in [2 \tau] \}$ of i.i.d. Bernoulli r.v.s
with $\Prob{}{\hat X_{i,t} = 1} = 1/d$, then $Y_t$ is stochastically smaller than 

\[ 
	\hat Y_t = \sum_{i = 1}^d \hat X_{i,t}  
\] 
For any node $u$ and any round $t$, the r.v. $\Q_t$ is thus 
stochastically smaller than the r.v. $\hat{\Q}_t$ defined recursively as follows.

\[ 
	\left\{
		\begin{array}{ccl}
			\hat{\Q}_t & = & \hat{\Q}_{t-1} + \hat Y_t - \chi_t \\
			\hat{\Q}_0 & = & 1 
		\end{array}
		\right. \ \ \mbox{ where } \ 
		\chi_t = \left\{
			\begin{array}{cc}
				1 & \mbox{ if } \hat{\Q}_{t-1} > 0 \\
				0 & \mbox{ otherwise }
			\end{array}
			\right.
		\]
Since our goal is to provide a concentration upper bound on $\Q_t$, we can do it by considering the 
		``simpler'' process $\hat \Q_t$. By the way, unrolling  $\hat \Q_t$ directly is far from trivial:
		 We   need
		the
		``right'' way to write it by using only i.i.d. Bernoulli r.v.s. Let's see how. 

		\noindent
		For any $t \in[2 \tau]$ and for any $s \in [t]$, define the r.v. 
		\begin{equation}\label{soda-eq:ardef}
			Z_{s,t} = \sum_{i = s}^t \hat Y_i - (t-s)
		\end{equation}

		\noindent
		Informally speaking, $Z_{s,t}$ matches the value of $\hat{\Q}_t$ whenever $s\leq t$ was the
		last previous round s.t. 
		$\hat{\Q}_s=0$. 

		As a key-fact 
		we show that $\hat{\Q}_t$ can be bounded by the maximum of $Z_{s,t}$ for $s \leq t$. 

		\begin{quote}
			\begin{claim}
				For any $t \in [2 \tau]$ it holds that
				\[
					\hat{\Q}_t \leqslant \max \{ Z_{s,t} \,:\, s = 1, \dots, t \} \ 
				\]
				and thus
				\begin{equation} \label{soda::boundqq}
					\max \{\Q_t \,:\, 1 \leqslant t \leqslant 2  \tau \} 
						\leqslant \max \{ Z_{s,t} \,:\, 1 \leqslant s \leqslant t \leqslant 2  \tau \} 
				\end{equation}
			\end{claim}

			\proof (of the Claim).
			For any $ s \in [t]$, let 
			\[ \chi_{s,t} = \prod_{r = s}^t \chi_r \] be the r.v. taking value $1$
			if $\hat{\Q}_{r-1} > 0$ for all $s \leqslant r \leqslant t$ and $0$ otherwise. It is easy
			 to prove by induction
			that $\hat{\Q}_t$ can be written as
			\begin{equation}\label{eq:qtrew}
				\hat{\Q}_t = 
				\sum_{s=2}^t (1 - \chi_{s-1}) \chi_{s,t} Z_{s-1, t} + \chi_{1,t} Z_{1,t} + (1 -\chi_t) Z_{t,t} 
			\end{equation}
			Since 
			$$
			\sum_{s=2}^t (1 - \chi_{s-1}) \chi_{s,t} + \chi_{1,t} = 1 
			$$
			the sum in  (\ref{eq:qtrew}) is not larger than the maximum of the $Z_{s,t}$, hence 

			\[ \hat{\Q}_t \leqslant \max \{ Z_{s,t} \,:\, s = 1, \dots, t \} \ \mbox{ and } \ 
			\max \{\Q_t \,:\, 1 \leqslant t \leqslant 2  \tau \} 
				\leqslant \max \{ Z_{s,t} \,:\, 1 \leqslant s \leqslant t \leqslant 2 \tau \} 
			\]
			\qed (of the Claim).
		\end{quote}

		\smallskip
		\noindent
		 Let us consider    (\ref{soda-eq:ardef}): 
		The r.v. $Z_{s,t} + (t-s)$ is a sum of $d \cdot (t - s +1)$ i.i.d. Bernoulli r.v.s each one with 
		expectation $1/d$. From the Chernoff bounds \eqref{eq:CB_upper_multip} and \eqref{eq:CB_absolute}, 
		for any $1 \leqslant s \leqslant t$,  it holds that
		$$
		\Prob{}{ Z_{s,t} \leqslant \max\left\{ \sqrt{c (t-s+1) \log n}, \, 6 c \log n \} \right\} } \geqslant 1 - n^{-c/3}
		$$
		By taking the union bound over all $1 \leqslant s \leqslant t \leqslant 2 \tau$,
		 from the above bound and  (\ref{soda::boundqq})
		we can get the desired 
		concentration bound on the maximal node congestion during every phase:

		$$
		\Prob{}{\max_{1 \leqslant t \leqslant 2\tau} \Q_t 
			\leqslant \max \left\{ \sqrt{ 2 c \tau\log n}, \, 6 c \log n \right\}} 
			\geqslant 1 - \frac{(2 \tau)^2}{n^{c/3}}
		$$

		\qed

		\smallskip

		As a consequence of the above Lemma, we can set the right value of 
		  $\tau$, thus getting the following result.

		\begin{theorem}
			Let $G = ([n],E)$ be a $d$-regular graph with $\tm{G}{1/4} = \polylog(n)$. 
			Each round of the \threestate\ on the clique can be simulated on $G$  in the \gossip\ model 
			 in $\polylog(n)$ rounds by exchanging messages of $\polylog(n)$ size, w.h.p.
		\end{theorem}
		\proof
		Let $2\tau = \alpha \bar{t}^2 \log n$ be the length of the phase, 
		where $\bar{t} = \tm{G}{1/n^2}$ and $\alpha$ is a suitable constant that we fix later.
		From Lemma~\ref{lemma:congestion}, we have that the maximum number of tokens in every node at any round
		of the phase is w.h.p at most
		$$
		\sqrt{2c\tau \log n} = \sqrt{\alpha c} \cdot \bar{t} \log n
		$$
		Since tokens are enqueued with a FIFO policy, 
		each single hop of the random walk performed by a token can be delayed 
		for at most the above number of rounds.
		Hence, in order to perform $\bar{t}$ hops of the random walk, 
		a token takes at most $\sqrt{\alpha c} \cdot \bar{t}^2 \log n$ rounds w.h.p. 

		\noindent
		By choosing $\alpha \geqslant 4 c$ we have that this number is smaller than $\tau$,  this 
		allows us to set   $\tau$ so that the forward process and the backward one 
		can both complete safely.

		\noindent
		By union bounding over all tokens we thus have that during the phase all tokens perform
		at least  $\bar{t}$ hops of a random walk and report back to the sender 
		the color of the node they reached after $\bar{t}$ hops w.h.p.

		\noindent
		Finally, notice that from  (\ref{eq:tvfall}) it follows that $\bar{t} = \polylog(n)$.  The phase length  
		and the size of the  exchanged messages are thus  $\polylog(n)$  as well. 

		\qed

		\smallskip
		Since a lazy random walk on regular expanders (see e.g.~\cite{HLW06}) has  $\polylog(n)$ mixing time, 
		from the above theorem and our result on the \threestate\ on the clique we easily get the following final result.

		\begin{cor}
			From any initial configuration $\bar{\genc}$ such that the \threestate\ 
			on the clique completes plurality consensus in $O(\mdist(\bar \genc) \log n)$ 
			rounds w.h.p., the modified \linebreak \threestate\ completes 
			plurality consensus on  any $d$-regular expander graph within  
			$O(\mdist(\bar \genc)  \cdot \polylog(n))$ rounds w.h.p.
		\end{cor}

%% file: trunk/conclusions.tex
\section{Open Problems}
There are several open  research directions related to the plurality problem on the gossip model.
One of the most interesting (and challenging)
ones concerns the monochromatic distance we have introduced in this paper.
We   believe   that this distance might represent a  general 
lower bound on the convergence time of  \emph{any} plurality dynamics which uses only 
$\log k + \Theta(1)$ bits of local memory.

Another interesting future research is the study of the \threestate\ (or other simple dynamics) over 
other classes of graphs. In our paper, we combined this dynamics with parallel random walks in order
to get an efficient protocol for regular expander graphs. We believe that similar  protocols can work
also in other classes of graphs such as \ErdRen\ graphs and dynamic graphs \cite{CMMPS08,CCDFIPPS13}.

%% file: trunk/appendix.tex
\section{Appendix} \label{appendix}

\begin{lemma}[Chernoff Bound, multiplicative form] \label{lem:cbmult}
Let $\left\{ X_{i}\right\} _{i\in\left[n\right]}$ be $n$ independent
r.v.s and let $\delta\in(0,1]$. It holds
\begin{align}
	\probb{\sum_{i\in\left[n\right]}X_{i}\leq\left(1-\delta\right)\cdot\mu_1}
		& \leq e^{-\frac{\delta^{2}}{2}\cdot\mathbb{E}\left[\sum_{i\in\left[n\right]}X_{i}\right]} 
		& \mbox{with \ensuremath{\mu_1 
		\leq \expe{\sum_{i\in\left[n\right]}X_{i}}}}
		\label{eq:CB_lower_multip}\\
	\probb{\sum_{i\in\left[n\right]}X_{i}\geq\left(1+\delta\right)\cdot\mu_2} 
		& \leq e^{-\frac{\delta^{2}}{3}\cdot\mathbb{E}\left[\sum_{i\in\left[n\right]}X_{i}\right]} 
		& \mbox{with \ensuremath{\mu_2 \geq \expe{\sum_{i\in\left[n\right]}X_{i}}} }
		\label{eq:CB_upper_multip}\\
	\probb{\sum_{i\in\left[n\right]}X_{i}\geq\mu_3 } 
		& \leq2^{- \mu_3} & \mbox{with \ensuremath{\mu_3 
		\geq6\cdot\expe{\sum_{i\in\left[n\right]}X_{i}}}}
		\label{eq:CB_absolute}
\end{align}
In particular, to obtain high probability, when $\expe{\sum_{i\in\left[n\right]}X_{i}}=\omega\left(\log n\right)$
in  (\ref{eq:CB_lower_multip}) and (\ref{eq:CB_upper_multip})
we can set $\delta=\sqrt{\frac{a\cdot\log n}{\expe{\sum_{i\in\left[n\right]}X_{i}}}}$
for any positive constant $a$.\end{lemma}

\begin{lemma}[Chernoff Bound, additive form]\label{lemma:cbadditive}
Let $X_1, \dots, X_n$ be a sequence of independent $\{0,1\}$ r.v.s, let $X = \sum_{i=1}^n X_i$ be their sum, and let $\mu = \expe{X}$. Then for $0 < \lambda < 1$ it holds that
\[
\Prob{}{X \geqslant \mu + n \lambda} \leqslant e^{- 2 n \lambda^2} 
\qquad \mbox{ and } \qquad 
\Prob{}{X \leqslant \mu - n \lambda} \leqslant e^{- 2 n \lambda^2}.
\]
\end{lemma}

	\begin{lemma}
		\label{lem:wop_intersection}Let $a$ and $b$ be two constants such that $a>b>0$,
		let $B$ be an event and let
		$\left\{ A_{i}\right\} _{i\in I}$
		be a family of events such that 
		$|I|=O\left( n^{b} \right)$ and $\probc{A_{i}}{B}\geq 1-n^{a}$. 
		Then, the event $\bigcap_{i\in I}``A_{i}|B\text{''}$
		holds with probability at least $1-\frac{|I|}{n^a}$.
	\end{lemma}
		\proof
			From the union bound
			\begin{align*}
				\probc{\bigcap_{i\in I}A_{i}}{B}=1-\probc{\bigcup_{i\in I}``\mbox{not }A_{i}\text{''}}{B}\geq 1-\frac{|I|}{n^a}
			\end{align*}
		\qed

	\begin{lemma}
		\label{lem:inequality}If $f\left(n\right)=\omega\left(1\right)$
		and $g\left(n\right)=o\left(f\left(n\right)\right)$ then 
		\[
			\left(1\pm \frac{1}{f\left(n\right)}\right)^{g(n)}
			=1\pm O\left(\frac{g\left(n\right)}{f\left(n\right)}\right)
		\]
	\end{lemma}
		\proof
			Use the elementary inequalities $e^{-\frac{x}{1-x}}\leq1-x\leq e^{-x}\leq1-\frac{x}{1+x}$
			for $|x|<1$.
		\qed

%% file: main.bbl
\begin{thebibliography}{10}

\bibitem{AD12}
M.~A. Abdullah and M.~Draief.
\newblock Majority consensus on random graphs of a given degree sequence.
\newblock {\em arXiv:1209.5025}, 2012.

\bibitem{AAKKLT08}
N.~Alon, C.~Avin, M.~Koucky, G.~Kozma, Z.~Lotker, and M.~R. Tuttle.
\newblock Many random walks are faster than one.
\newblock In {\em Proc. of 20th ACM SPAA}, 2008.

\bibitem{AAE07}
D.~Angluin, J.~Aspnes, and D.~Eisenstat.
\newblock A simple population protocol for fast robust approximate majority.
\newblock {\em Distributed Computing}, 21(2):87--102, 2008.
\newblock Ext. abs. in DISC 2007.

\bibitem{BD13}
A.~Babaee and M.~Draief.
\newblock Distributed multivalued consensus.
\newblock In {\em Computer and Information Sciences III}, pages 271--279.
  Springer, 2013.

\bibitem{BCNPST13}
L.~Becchetti, A.~Clementi, E.~Natale, F.~Pasquale, R.~Silvestri, and
  L.~Trevisan.
\newblock Simple dynamics for plurality consensus.
\newblock In {\em Proc. of 26th ACM SPAA}, 2014.

\bibitem{BTV09}
F.~B{\'e}n{\'e}zit, P.~Thiran, and M.~Vetterli.
\newblock Interval consensus: from quantized gossip to voting.
\newblock In {\em Proc. of 34th IEEE ICASSP}, 2009.

\bibitem{BGPS06}
S.~Boyd, A~Ghosh, B.~Prabhakar, and D.~Shah.
\newblock Randomized gossip algorithms.
\newblock {\em IEEE Transactions on Information Theory}, pages 2508--2530,
  2006.

\bibitem{cardelli2012cell}
L.~Cardelli and A.~Csik{\'a}sz-Nagy.
\newblock The cell cycle switch computes approximate majority.
\newblock {\em Scientific Reports}, Vol. 2, 2012.

\bibitem{CHKM12}
K.~Censor-Hillel, B.~Haeupler, J.~Kelner, and P.~Maymounkov.
\newblock Global computation in a poorly connected world: Fast rumor spreading
  with no dependence on conductance.
\newblock In {\em Proc. of 44th ACM STOC}, 2012.

\bibitem{CCDFIPPS13}
A.~Clementi, P.~Crescenzi, C.~Doerr, P.~Fraigniaud, M.~Isopi, A.~Panconesi,
  F.~Pasquale, and R.~Silvestri.
\newblock Rumor spreading in random evolving graphs.
\newblock In {\em Proc. of 21st ESA}, 2013.

\bibitem{CDGNS13}
A.~Clementi, M.~Di~Ianni, G.~Gambosi, E.~Natale, and R.~Silvestri.
\newblock Distributed community detection in dynamic graphs.
\newblock In {\em Proc. of 20th SIROCCO}, 2013.

\bibitem{CMMPS08}
A.~Clementi, C.~Macci, A.~Monti, F.~Pasquale, and R.~Silvestri.
\newblock Flooding time of edge-markovian evolving graphs.
\newblock {\em SIAM J. Discrete Math.}, 24(4):1694--1712, 2010.

\bibitem{CER14}
C.~Cooper, Elsasser R., and Tomasz Radzik.
\newblock The power of two choices in distributed voting.
\newblock In {\em Proc. of 41st ICALP}, 2014.

\bibitem{SMP12}
A.~Das~Sarma, A.R. Molla, and G.~Pandurangan.
\newblock Near-optimal random walk sampling in distributed networks.
\newblock In {\em Proc. of 31st IEEE INFOCOM}, 2012.

\bibitem{DGHILSSST87}
A.~Demers, D.~Greene, C.~Hauser, W.~Irish, J.~Larson, S.~Shenker, H.~Sturgis,
  D.~Swinehart, and D.~Terry.
\newblock {Epidemic algorithms for replicated database maintenance}.
\newblock In {\em Proc. of 6th ACM PODC}, 1987.

\bibitem{DGMSS11}
B.~Doerr, L.~A. Goldberg, L.~Minder, T.~Sauerwald, and C.~Scheideler.
\newblock Stabilizing consensus with the power of two choices.
\newblock In {\em Proc. of 23rd ACM SPAA}, 2011.

\bibitem{Doty14}
D.~Doty.
\newblock Timing in chemical reaction networks.
\newblock In {\em Proc. of 25th ACM-SIAM SODA}, 2014.

\bibitem{DV12}
M.~Draief and M.~Vojnovic.
\newblock Convergence speed of binary interval consensus.
\newblock {\em SIAM J. on Control and Optimization}, 50(3):1087--1109, 2012.

\bibitem{HPPRS12}
B.~Haeupler, G.~Pandurangan, D.~Peleg, R.~Rajaraman, and Z.~Sun.
\newblock Discovery through gossip.
\newblock In {\em Proc. of 24th ACM SPAA}. ACM, 2012.

\bibitem{HLW06}
S.~Hoory, Linial N., and Wigderson W.
\newblock Expander graphs and their applications.
\newblock {\em Bull. Amer. Math. Soc. (N.S)}, 43:439--561, 2006.

\bibitem{KDG03}
D.~Kempe, A.~Dobra, and J.~Gehrke.
\newblock Gossip-based computation of aggregate information.
\newblock In {\em Proc. of 43rd IEEE FOCS}, 2003.

\bibitem{lpwAMS08}
D.~Levin, Y.~Peres, and E.~L. Wilmer.
\newblock {\em Markov Chains and Mixing Times}.
\newblock AMS, 2008.

\bibitem{Spir14}
G.~B. Mertzios, S.~E. Nikoletseas, C.~Raptopoulos, and P.~G. Spirakis.
\newblock Determining majority in networks with local interactions and very
  small local memory.
\newblock In {\em Proc. of 41st ICALP}, 2014.

\bibitem{MNT12}
E.~Mossel, J.~Neeman, and O.~Tamuz.
\newblock Majority dynamics and aggregation of information in social networks.
\newblock {\em Autonomous Agents and Multi-Agent Systems}, pages 1--22, 2012.

\bibitem{MS10}
E.~Mossel and G.~Schoenebeck.
\newblock Reaching consensus on social networks.
\newblock In {\em Proc. of 2nd ITCS}, 2010.

\bibitem{MNT14}
Elchanan Mossel, Joe Neeman, and Omer Tamuz.
\newblock Majority dynamics and aggregation of information in social networks.
\newblock {\em Autonomous Agents and Multi-Agent Systems}, 28(3), 2014.

\bibitem{FKP11}
Fraigniaud P., Korman A., and Peleg D.
\newblock Local distributed decision.
\newblock In {\em Proc. of 52nd IEEE FOCS}, 2011.

\bibitem{P00}
D.~Peleg.
\newblock {\em Distributed Computing: A Locality-Sensitive Approach}.
\newblock SIAM, 2000.

\bibitem{Pe02}
D.~Peleg.
\newblock Local majorities, coalitions and monopolies in graphs: a review.
\newblock {\em Theor. Comput. Sci.}, 282(2):231--257, 2002.

\bibitem{PVV09}
E.~Perron, D.~Vasudevan, and M.~Vojnovic.
\newblock Using three states for binary consensus on complete graphs.
\newblock In {\em Proc. of 28th IEEE INFOCOM}, 2009.

\bibitem{RM08}
Y.~Ruan and Y.~Mostofi.
\newblock Binary consensus with soft information processing in cooperative
  networks.
\newblock In {\em Proc. of 47th IEEE CDC}, 2008.

\end{thebibliography}
